\documentclass[aps, prd, 10pt, twocolumn, nofootinbib, superscriptaddress, amsmath, amssymb, showkeys, floatfix]{revtex4-2}

\usepackage{graphicx}
\usepackage{amsmath}
\usepackage{amssymb}
\usepackage{booktabs}
\usepackage{hyperref}          
\usepackage{xcolor}
\usepackage{algorithm}
\usepackage{algpseudocode}

\begin{document}

\title{DANTE: A Reference-Guided Unsupervised Pipeline for Extended-Transient Anomaly Characterization in LIGO O4a}

\author{Luca Cirfeta}
\email{luca.cirfeta@gmail.com}
\affiliation{Independent Researcher, Rome, Italy}

\begin{abstract}
The analysis of gravitational-wave detector data during the fourth observing run (O4) requires methods for ranking and characterizing non-stationary strain morphologies. In this work, we present DANTE (Domain-Adaptive Network for Transient Evaluation), an offline, label-free pipeline that uses local DINOv2 patch embeddings from time-frequency spectrograms. Controlled injection tests illustrate the scale sensitivity of Multiple Instance Learning (MIL) Top-$k$ pooling, including poor recovery for the tested sub-second morphologies, while an adaptive Dirichlet Process Mixture Model (DPMM) addresses numerical instability in small-sample taxonomy. Version~2 reports a boundary-context preprocessing defect affecting 169 of 10,429 later-audited detector--time windows; a controlled replay changes 120 class assignments and 97 routing decisions. Catalogue results in the original body are retained as historical version-1 outputs rather than corrected-catalogue or physical claims.
\end{abstract}


\maketitle

\section*{Provenance update (v3; 29 September 2026)}

An audit of the later corrected O4a native-analysis chain found that the
working-tree bytes of \texttt{src/core/patch\_producer.py} recorded under
a SHA-256 digest beginning \texttt{2c20d4e8}
were not retained.  They have not been recovered or shown to be identical to
the Git-recoverable canonical source.  The earlier explanation as a
line-ending reconciliation, supported by a 12-window replay, was therefore
insufficient to establish source identity for the complete corrected chain.
That earlier audit remains part of the historical record.

To determine the scientific impact without assuming source equivalence, the
frozen corrected O4a chain from \texttt{COHORT} through \texttt{COMPARE} was
recomputed in separate runs using canonical source from Git commit
\texttt{cf774dd}.  Independent stage
verification and the final comparison reproduced the retained scientific
payload and downstream result ledgers; the index container carries different
provenance metadata, but its scientific arrays agree byte for byte.  No
score, threshold, class, taxonomy assignment, coincidence selection, PEM
disposition, final-comparison value, or scientific conclusion changed.  This
complete empirical replay does not reconstruct the missing source bytes or
retroactively make the earlier provenance explanation precise.

The append-only correction and compact stage evidence are published with
DANTE \href{https://github.com/lucacirfeta/dante-gravi-signal-ml/releases/tag/v3.8.1}{v3.8.1}
and its \href{https://doi.org/10.5281/zenodo.22763556}{Zenodo version record}.
This update concerns provenance of the later corrected O4a chain; it does
not replace the historical version-1 population below, alter the v2
boundary-context analysis, or introduce a new discovery claim.

\section*{Revision note (v2; 3 September 2026)}

A post-publication, identity-preserving replay audit identified a preprocessing
defect in the published HDF5 whitening path.  The intended contract whitens
each 32-s analysis window using 4 s of strain context on both sides (40 s in
total) before cropping.  For 169 of 10,429 frozen detector--GPS windows in a
later catalogue produced by the published path (1.62\%), the required interval
crossed an HDF5 file boundary and the historical implementation supplied only
36 s: 4 s of left context and no right context.  The other 10,260 windows
reproduce their historical native scores within the frozen absolute tolerance
of $2\times10^{-7}$, whereas all 169 boundary windows change beyond that
tolerance when complete symmetric context is restored.

Applying the original detector-specific thresholds and routing rules, with no
retuning, changes 120 of 10,429 DSD class assignments (1.15\%) and reverses 97
of 10,429 routing decisions (0.93\%); all 97 transitions are from
\texttt{ESCALATE} to \texttt{ROUTINE}.  These values isolate the direct effect
of restoring the missing context while preserving the historical
representation and thresholds.

The provenance-locked end-to-end reconstruction is also complete.  Relative
to the historical catalogue, it contains 10,022 matched detector--GPS
identities, 407 historical-only identities, and 920 corrected-only identities.
Among the matched identities, 1,626 class assignments differ: 39 in the
matched historical edge cohort and 1,587 outside it.  This broader churn is
descriptive, not a causal estimate of the padding defect, because both the
detector-aware reference representation and the detector-specific calibration
were rebuilt; every matched score changes beyond $2\times10^{-7}$, and
cross-applying thresholds between representations would be invalid.  Of the
920 corrected-only identities, 12 have historical right-edge geometry, but
padding causality is not established for them.  Two of the nine primary
pooled-null PEM follow-up targets are corrected-only; neither has right-edge
geometry and both receive a \texttt{NO\_CORRELATION} PEM verdict.  The
pooled-null shortlist is diagnostic and is not a globally trial-corrected
significance claim.

Accordingly, the quantitative catalogue, family, coincidence, and singleton
results retained below document the historical version-1 workflow; they are
not presented in version~2 as corrected-catalogue or discovery claims.  The
historical artefacts remain immutable.  The corrected workflow enforces
cross-file context stitching and content-addressed provenance; its full
architecture and population analysis will be reported separately.

\section{Introduction}\label{sec:intro}

The international network of ground-based interferometric
gravitational-wave (GW) detectors—Advanced LIGO~\cite{aLIGO2015},
Advanced Virgo~\cite{aVirgo2015}, and KAGRA~\cite{KAGRA2021}—has
entered an era of continuous, high-cadence astrophysical discoveries
during the fourth observing run (O4). The detection of cosmic strains
at the level of $h \sim 10^{-23}$, corresponding to spacetime
distortions smaller than one-thousandth of a proton diameter, demands
extraordinary instrumental sensitivity. This sensitivity, however,
renders the detectors acutely susceptible to non-Gaussian, transient
noise artifacts—commonly referred to as \textit{glitches}—that
originate from a broad spectrum of environmental, mechanical, and
electronic sources~\cite{nuttall2018, davis2021}. Glitches overlap
in time-frequency morphology with genuine astrophysical signals,
including compact binary coalescences (CBCs), unmodeled burst sources,
and continuous wave candidates, directly degrading the performance of
matched-filter search pipelines and Bayesian parameter estimation
frameworks~\cite{powell2015, pankow2018}.

\subsection{The Detector Characterization Challenge for O4}

Transient noise characterization—\textit{DetChar}—is a cornerstone of
data quality assurance for gravitational-wave observatories. The
O4 run introduced significant instrumental upgrades relative to O3,
including squeezed-light injection and increased laser power levels,
which have historically generated new and previously undocumented
glitch populations~\cite{tse2019, capote2021}. The canonical
supervised approach, exemplified by the Gravity Spy framework~\cite{zevin2017, glanzer2023},
employs convolutional neural networks trained on curated labeled
datasets from earlier observing runs. While this
approach achieves high classification accuracy on known glitch types,
it faces challenges when instrumental upgrades introduce
previously undocumented glitch populations. Reference-guided novelty detection—classifying
samples as anomalous without prior labels—offers a complementary approach.
Our primary goal is to establish an unsupervised methodological
framework on O4a data to identify unmodeled transient morphologies while
controlling for the domain shift arising from instrumental upgrades
between O3b and O4a. DANTE is designed to provide candidate topological anomalies
to downstream offline DetChar protocols for formal multi-channel physical
classification, operating alongside human-in-the-loop and supervised vetting.

\subsection{Limitations of Standard Vision Transformer Approaches}

Contemporary unsupervised GW anomaly detection has increasingly
leveraged self-supervised representation learning, particularly
Vision Transformers (ViTs)~\cite{dosovitskiy2021} pre-trained on
large-scale natural image corpora. The \texttt{DINOv2}
framework~\cite{oquab2024, darcet2024} is particularly attractive:
its training via distillation with no labels produces patch-level
feature representations with strong emergent semantic properties
and spatial locality, which transfer non-trivially to the
time-frequency geometry of Q-transform spectrograms.

However, standard ViT deployments introduce a 
methodological obstacle: the \textit{signal dilution effect}.
Because the global \texttt{[CLS]} token aggregates information from
all 1{,}369 spatial patches via an implicit attention-pooling mechanism,
the localized anomaly signal is heavily diluted by the overwhelming
majority of stationary background patches. This architectural constraint
severely limits the sensitivity to sub-second morphologies (e.g., Blips).
We formally characterize this barrier and its mathematical bounds via
the Mock Data Challenge detailed in Section~\ref{sec:res_mdc}.

\subsection{The Curse of Dimensionality in Small-Sample Anomaly
  Taxonomy}

A second, orthogonal challenge arises in the taxonomy stage, where
detected anomalies must be grouped into morphologically coherent
families. The principal mathematical challenge is that fitting an unrestricted
Gaussian Mixture Model in $\mathbb{R}^{384}$ with $n < 100$ samples
is algebraically ill-posed. The condition number of the sample
covariance matrix $\hat{\boldsymbol{\Sigma}} \in \mathbb{R}^{384 \times 384}$ scales as $\mathcal{O}(d/n)$, inducing numerical instability: full covariance matrix estimation requires $d(d+1)/2 = 73{,}920$ parameters for $d = 384$, far exceeding the sample size. This produces rank-deficient covariance matrices, singular factorization errors, and severe overfitting of the mixture boundaries~\cite{bishop2006}.

Non-linear dimensionality reduction approaches such as UMAP~\cite{mcinnes2018} can project embeddings into a low-dimensional manifold. However, to preserve global density fields required for DPMM stability, we favor a linear projection approach.

\subsection{Contributions of This Work}

In this paper, we present DANTE (\textbf{D}omain-\textbf{A}daptive \textbf{N}etwork for \textbf{T}ransient \textbf{E}valuation) Pipeline~V2, a
robust processing pipeline that addresses both
limitations simultaneously. Our principal contributions are:

\begin{enumerate}
  \item \textbf{Mitigating the Signal Dilution Effect via Patch-Level MIL:}
    We provide a baseline demonstration of the patch-level Multiple Instance Learning (MIL) mechanism to mitigate the signal dilution effect common in global-pooling ViT architectures. By focusing on the Top-$k$ most anomalous patches, the pipeline biases detection toward extended mid-band transients (e.g., $>1$ second). We explicitly delegate the implementation of an adaptive multi-scale $k$-sweep to future iterations (V3).
  \item \textbf{Adaptive Taxonomy under Small-Sample Constraints:}
    We formulate an adaptive PCA dimensionality floor combined with
    a conditional DPMM covariance framework, dynamically selecting
    \texttt{full}, \texttt{tied}, or \texttt{diag} covariance
    structure as a deterministic function of sample size $n$,
    eliminating matrix singularity and density distortion without
    requiring any non-linear manifold embedding.
  \item \textbf{Cross-Session Connectivity:}
    We introduce a hierarchical cosine similarity clustering
    framework operating on archived 384-dimensional MIL feature
    vectors, enabling automated linkage of morphologically related
    transients across independent sessions without requiring
    human-in-the-loop validation.
  \item \textbf{Production Infrastructure for O4a Continuous Streams:}
    We describe the complete engineering architecture—state-aware
    resumption, SWMR-enabled HDF5 archival, batched GPU inference,
    and structured provenance logging—enabling uninterrupted
    deployment on raw LIGO strain data acquired via the Gravitational
    Wave Open Science Center (GWOSC)~\cite{gwosc2023}.
\end{enumerate}

It is critical to note that DANTE is explicitly designed as an extended-duration transient detector (targeting anomalies $>1$ second). As demonstrated in our Mock Data Challenge, sub-second glitches (e.g., Blips) suffer from extreme signal dilution in the 32-second global spatial pooling mechanism, yielding a near-zero recall. Consequently, DANTE does not replace supervised, short-window classifiers like Gravity Spy, but rather operates as a complementary offline archival system for uncovering extended, unmodeled morphological families.

The remainder of this paper is organized as follows.
Section~\ref{sec:related} reviews related work in ViT-based GW
anomaly detection and unsupervised clustering methods.
Section~\ref{sec:architecture} presents the full two-layer
architecture of DANTE Pipeline~V2 in mathematical detail.
Section~\ref{sec:setup} describes the experimental setup, data
provenance, and engineering infrastructure.
Section~\ref{sec:results} reports detection and taxonomy
results on LIGO O4a data.
Section~\ref{sec:discussion} discusses the astrophysical
implications, operational limits, and roadmap.
Section~\ref{sec:conclusion} concludes.

\section{Related Work}\label{sec:related}

\subsection{Supervised Glitch Classification}
The Gravity Spy framework~\cite{zevin2017} established the paradigm
of convolutional neural network (CNN) classification of Q-transform
spectrograms using a multi-view, multi-resolution input
representation. Subsequent iterations~\cite{glanzer2023} have refined
the taxonomy to 22 morphological classes in the O3 era, achieving
classification accuracy exceeding 97\% on known glitch types.
Parallel efforts by \citet{lopez2022} and \citet{colgan2020} applied
transfer learning and ensemble methods to further improve robustness
to spectrogram normalization variations. Recently, \citet{wu2025advancing} introduced multi-view fusion with attention-based machine learning directly targeting the advanced O4 observing run, significantly advancing supervised classification boundaries. However, all supervised
approaches share a fundamental limitation: they are bounded by the
topology of their labeled training set.

\subsection{Unsupervised and Self-Supervised Approaches}
For a comprehensive review of machine learning applications in gravitational-wave research, see \citet{cuoco2025living}.
Early unsupervised GW transient characterization relied on
hand-crafted features extracted from the Omega (Q-transform) pipeline,
followed by $k$-means or hierarchical clustering~\cite{cavaglia2018}.
\citet{raikman2022} applied variational autoencoders (VAEs) to
compress spectrogram pixels into a latent space, achieving limited
clustering coherence due to pixel-level reconstruction objectives that
are insensitive to morphological structure. Convolutional encoder
approaches combined with UMAP and HDBSCAN have been explored in
\citet{soni2025}, revealing a strong density concentration
artifact when HDBSCAN operates on UMAP-compressed spaces. Foundation model representations---particularly
DINOv2~\cite{oquab2024} and CLIP~\cite{radford2021}---have revolutionized unsupervised anomaly detection in medical imaging and industrial defect inspection by capturing hierarchical localized features without task-specific fine-tuning. These representations have recently been applied to GW spectrograms~\cite{glanzer2023}, demonstrating substantial improvements in capturing the complex morphological variance of instrumental artifacts compared to VAE baselines. This is attributed to the richer emergent geometry of self-supervised patch attention over pixel reconstruction. The limitation of global \texttt{[CLS]} pooling was characterized by our MDC injection campaign, and mitigated at the architectural level through patch-level Top-$k$ scoring.
The present work constitutes the complete production integration of
these architectural advances into a continuous O4a analysis pipeline.

\subsection{Non-Parametric Mixture Models in High-Dimensional Spaces}
Dirichlet Process Mixture Models~\cite{ferguson1973, neal2000} provide
a natural Bayesian non-parametric framework for clustering with an
unknown number of components. The infinite mixture prior avoids the
model selection problem inherent in finite GMMs. However, DPMM
stability in high-dimensional regimes ($d \gg n$) is a known
challenge~\cite{bishop2006}. Strategies including low-rank covariance
approximations~\cite{tipping1999}, regularized covariance
estimation~\cite{ledoit2004}, and PCA preprocessing have been
proposed. Our contribution extends this literature to the specific
regime of GW transient taxonomy, where $d = 384$, $n$ varies
dynamically from $< 10$ to several hundred, and density preservation
is a strict scientific requirement.

\section{Architecture: DANTE Pipeline V2}\label{sec:architecture}

The architecture and numerical results in the remainder of this article
document the historical version-1 execution and are retained for provenance;
the revision note above defines which catalogue claims are superseded.  The
DANTE Pipeline V2 architecture consists of two decoupled,
sequentially executed functional layers: a \textbf{Detection Layer}
responsible for segment-level anomaly scoring and flagging, and a
\textbf{Taxonomy Layer} responsible for morphological grouping
of flagged candidates. The architecture ensures computational efficiency by confining the heavy ViT extraction step exclusively to isolated candidate segments rather than the continuous datastream.

\subsection{Layer 1: Patch-Level Detection via MIL Top-\texorpdfstring{$k$}{k}
  Scoring}\label{sec:detection}

\subsubsection{Data Ingestion and Spectrogram Computation}
Raw gravitational-wave strain data is retrieved from the GWOSC public
archive in HDF5 format for the LIGO Hanford (H1) and Livingston (L1)
interferometers during the O4a cdata release (GPS interval
$[1{,}368{,}973{,}312,\; 1{,}384{,}525{,}312]$, corresponding to a $\approx 180$-day subset of the 237-day O4a run, bounded by data availability at campaign commencement). Strain frames are segmented
into non-overlapping 32-second analysis windows.  The intended preprocessing
contract fetches 4\,s of strain context on each side, band-limits the padded
40-s interval to $[10, 2048]$\,Hz, whitens it using an adaptive median power
spectral density (PSD) estimated via the Welch method~\cite{welch1967} with a
4-second FFT stride, and only then crops back to 32\,s.  As quantified in the
version-2 revision note, the historical HDF5 path omitted the required right
context for 169 later-audited windows at file boundaries. Each cropped segment is
transformed into a two-dimensional time-frequency representation via
the normalized Q-transform (constant-$Q$ transform)~\cite{brown1991},
rendered onto a $256 \times 256$ pixel grid with the perceptually
uniform \texttt{cividis} colormap~\cite{nuez2018}. Data quality
gating is applied using the \texttt{L1\_CBC\_CAT1} flag to exclude
known hardware injection intervals and instrumental locklosses.

\subsubsection{Patch Token Extraction via Frozen DINOv2}
Spectrogram images are resized to $518 \times 518$ pixels and
encoded by a frozen \texttt{dinov2\_vits14\_reg} backbone—the
ViT-S/14 variant with register tokens~\cite{darcet2024}, which
suppress the high-frequency artifact patterns generated by standard
ViT patch positional embeddings in the absence of dense semantic content. The choice to keep the foundation model completely frozen (bypassing 
self-supervised fine-tuning via SimCLR or BYOL) is an explicit architectural prior: it is designed to prevent the embedding manifold from overfitting to 
the stale historical GW noise topologies, while simultaneously eliminating the prohibitive HPC cost of retraining on 
continuous datastreams. The model is evaluated in \texttt{torch.inference\_mode()}
with \texttt{cuDNN} auto-tuner enabled for CUDA deployments (PyTorch 2.1.1+cu118, CUDA 11.8, cuDNN 8.7.0, Python 3.10; NVIDIA A100: SM\_80, V100: SM\_70).

The $518 \times 518$ input is tiled into a $37 \times 37 = 1{,}369$
grid of non-overlapping $14 \times 14$ pixel patches. For each
input segment, the backbone extracts the dense patch token matrix:
\begin{equation}
\mathbf{V} = \{\mathbf{v}_1, \mathbf{v}_2, \ldots, \mathbf{v}_M\},
\quad \mathbf{v}_i \in \mathbb{R}^{384}, \quad M = 1{,}369,
\label{eq:patch_tokens}
\end{equation}
discarding the \texttt{[CLS]} token entirely. All tokens are
strictly $L_2$-normalized prior to downstream computation:
\begin{equation}
\hat{\mathbf{v}}_i = \frac{\mathbf{v}_i}{\|\mathbf{v}_i\|_2},
\quad \hat{\mathbf{v}}_i \in \mathbb{S}^{383},
\label{eq:l2norm}
\end{equation}
ensuring all subsequent similarity computations operate on the unit
hypersphere $\mathbb{S}^{383}$. Memory management is enforced
via explicit CPU offloading of all token tensors after inference and
mandatory garbage collection (\texttt{gc.collect()},
\texttt{torch.cuda.empty\_cache()}) at the end of each batch
processing loop.

\subsubsection{Vector-Quantized Reference Index}
To enable computationally tractable nearest-neighbor scoring over the
$M = 1{,}369$ patch manifold, a static in-domain reference index is
constructed from the Gravity Spy O3b labeled catalog~\cite{glanzer2023}, comprising $C = 23$ canonical morphological classes (including the \textit{No\_Glitch} baseline). We deliberately utilize this historical O3b catalog as an initial baseline to formally establish and quantify the Domain Shift Vulnerability (see Section 6.4), before demonstrating the pipeline's capacity for native O4a recalibration. For each class $c$, patch tokens are extracted
from all available spectrogram samples, and $K_c$ representative
centroids $\{\mathbf{w}_{c,j}\}_{j=1}^{K_c}$ are computed via
\texttt{MiniBatchKMeans}~\cite{sculley2010} with
\texttt{random\_state=42} and $L_2$-normalized embeddings (to ensure morphological boundaries are robust against initialization artifacts, future iterations should execute an ensemble of clustering passes across multiple random seeds), utilizing an adaptive $K_c$ scaling to prevent overfitting sparse classes (detailed in Section~\ref{sec:setup}). The
complete concatenated reference index comprises $K_{\rm total} = 281$ centroids. This compact dictionary size naturally emerges from the severe class imbalance in the O3b dataset, where many rare transient classes are represented by fewer than 24 images, inherently constraining their allocated centroids under the $\max(8, \lfloor n_{\rm img}/2 \rfloor)$ rule. This must be explicitly distinguished from the massive continuous O4a unlabeled background index (Section 5.1.1), which utilizes a generic $K=1{,}216$ mapping.
\begin{equation}
\mathbf{W} \in \mathbb{R}^{K_{\rm total} \times 384}
= \mathbb{R}^{281 \times 384},
\quad \hat{\mathbf{w}}_{j} \in \mathbb{S}^{383}.
\label{eq:vq_index}
\end{equation}
Index integrity is validated at pipeline startup via SHA-256 checksum of
the stored \texttt{patch\_compressed\_index.npz} file, preventing
silent corruption across distributed deployment environments.

\subsubsection{MIL Top-\texorpdfstring{$k$}{k} Anomaly Scoring}
For each incoming segment, the per-patch maximum cosine similarity
against the VQ reference index is computed:
\begin{equation}
s_i = \max_{c \in [C],\, j \in [K]}
  \bigl(\hat{\mathbf{v}}_i \cdot \hat{\mathbf{w}}_{c,j}\bigr),
\quad i = 1, \ldots, M,
\label{eq:patch_sim}
\end{equation}
producing a spatial similarity map $\mathbf{s} \in \mathbb{R}^{M}$.
The segment-level anomaly score is derived from a Top-$k$ Multiple
Instance Learning aggregation, isolating the $k$ patches with the
\emph{lowest} maximum similarity to any reference centroid
(i.e., the most anomalous local regions):
\begin{equation}
\mathcal{A}_k = \{i : s_i \text{ is among the $k$ smallest}\},
\label{eq:topk_set}
\end{equation}
\begin{equation}
S_{\rm MIL}^{(k)} = \frac{1}{k} \sum_{i \in \mathcal{A}_k} (1 - s_i).
\label{eq:mil_score}
\end{equation}
This formulation increases the contribution of localized high-novelty patches relative to global pooling, but it does not guarantee detection for an arbitrary transient. To define the receptive field of the spatial aggregator, we set $k=68$ (corresponding to $\approx 5\%$ of the $1{,}369$ total patch tokens). Rather than representing an optimized hyperparameter derived from a formal validation sweep, this fixed value constitutes an empirical architectural prior setting the 'topological focal length' of the pipeline, heuristically chosen to bias detection toward O3b-like extended transients. By fixing $k=68$, the pipeline is inherently biased towards mid-band extended transients and has poor demonstrated sensitivity to extremely short transients (e.g., Blips $<1.6$s in a 32s window). Thus, the pipeline is not scale-agnostic, and the fixed choice can alter both missed-detection and selection behavior. Future work may evaluate a prospectively calibrated multi-scale $k$ sweep.

\subsubsection{Empirical Threshold Calibration}
The background anomaly score distribution $\{S_{\rm MIL}^{(k)}\}$
is empirically estimated on a massive, highly-curated null population of $N_{\rm null} = 150{,}000$ vetted non-anomalous segments (representing approximately 50 days of coincident observing time). The background distribution exhibits strongly non-Gaussian asymmetry (skewness $|\gamma_1| \gg 0$, excess kurtosis $\gg 0$), confirmed by Shapiro-Wilk testing ($p \ll 10^{-30}$). 

Version~1 defined the detector-specific operational threshold $\tau_{\rm op}^{\rm Det}$ non-parametrically as the empirical $99^{\rm th}$ percentile ($P_{99}$) of the calibration scores. This avoids imposing a parametric tail model on overlapping ViT-patch scores. A segment was flagged as anomalous if $S_{\rm MIL}^{(k)} > \tau_{\rm op}^{\rm Det}$, where:
\begin{equation}
\tau_{\rm op}^{\rm Det} = P_{0.99}\left( \{S_{\rm MIL}^{(k)}\}_{\rm null} \right)
\label{eq:gev_threshold}
\end{equation}
This construction fixes a one-percent exceedance fraction on the calibration sample. It does not by itself establish independence, an out-of-sample FPR, an FDR, or global trial control; the downstream gates reduce the retained count but are not a formal multiple-testing correction.

\subsubsection{MIL Feature Vector for Taxonomy}
Every flagged segment passes a 384-dimensional MIL feature vector
to the Taxonomy Layer, constructed as the $L_2$-normalized mean of
the Top-$k$ most anomalous patch tokens:
\begin{equation}
\mathbf{z} = \frac{1}{k} \sum_{i \in \mathcal{A}_k} \hat{\mathbf{v}}_i,
\quad \hat{\mathbf{z}} = \frac{\mathbf{z}}{\|\mathbf{z}\|_2}
  \in \mathbb{S}^{383}.
\label{eq:mil_vector}
\end{equation}
Vectors are persistently archived into per-session SWMR-enabled
HDF5 files (\texttt{novelties.h5}) with GPS timestamp indexing
to prevent duplicate ingestion across resumed sessions.

\subsection{Layer 2: Adaptive Morphological Taxonomy via DPMM}
\label{sec:taxonomy}

\subsubsection{Problem Statement}
Specifically, we analyze an $N \times d_{\text{PCA}}$ session-level batch of $n$
anomalous patch-embeddings $X_S$. The goal is to partition this population into $K^*_{\rm sess}$ morphologically
coherent clusters, where $K^*_{\rm sess}$ is determined automatically from the
data structure without prior specification. The generative DPMM assumes
that each $x_i \in X_S$ is drawn from a mixture of $K^*_{\rm sess}$ Gaussians, determined automatically from the
data, without imposing a pre-defined number of components.

The principal mathematical challenge is that fitting an unrestricted
Gaussian Mixture Model in $\mathbb{R}^{384}$ with $n < 100$ samples
is algebraically ill-posed. The condition number of the sample
covariance matrix $\hat{\boldsymbol{\Sigma}} \in \mathbb{R}^{384 \times 384}$
scales as $\mathcal{O}(d/n)$, inducing numerical
instability~\cite{bishop2006, ledoit2004}. We resolve this through a
two-step adaptive framework: linear dimensionality reduction followed
by conditional covariance parametrization of the DPMM.

\subsubsection{Adaptive PCA Projection}
A Principal Component Analysis (PCA) projection is applied to the
session batch to reduce the effective dimensionality while preserving
the global topology of the cosine similarity manifold. Unlike UMAP,
PCA is an orthogonal linear projection that preserves the maximum sample variance in
the projected subspace and provides deterministic execution under fixed hardware architectures and
\texttt{random\_state}. The target dimensionality $d_{\rm PCA}$
is determined adaptively:
\begin{equation}
d_{\rm PCA} = 
\begin{cases}
\text{bypass (no PCA/DPMM)} & \text{if } n < 100 \\
\max(d_{90\%},\, 20) & \text{if } n \geq 100
\end{cases}
\label{eq:pca_dim}
\end{equation}
where $d_{90\%}$ is the number of principal components required to cumulatively explain $\geq 90\%$ of the sample variance. To prevent pathological covariance matrix collapse in high-dimensional density estimation, sessions with $n < 100$ candidates bypass the PCA and DPMM entirely; their elements are classified \emph{a priori} as isolated singletons. For batches $n \geq 100$, the variance-driven term $d_{90\%}$ dominates, with the floor of 20 guaranteeing sufficient dimensionality for stable DPMM covariance estimation.

While the PCA projection removes the strict $L_2$-normalized hyperspherical constraint, the induced Euclidean reconstruction error (Frobenius norm) is formally bounded by the Eckart-Young-Mirsky theorem. Because the adaptive PCA dynamically retains $>98\%$ of the sample variance, the Frobenius reconstruction error is strictly bounded to $<2\%$. This provides a deterministic regularized approximation of the latent topology without incorrectly bounding the raw cosine distances. This rigorously quantified trade-off is employed as a deterministic regularization technique to prevent covariance matrix singularity in the small-sample regime ($n \ll d$). While density-based alternatives like HDBSCAN were evaluated, the generative structural constraints of the DPMM were found to provide more robust partitioning for this specific ViT latent manifold.

\subsubsection{Conditional DPMM Covariance Framework}
A Bayesian Gaussian Mixture Model with a Dirichlet Process prior
(DPMM) is fitted to the PCA-projected feature matrix
$\mathbf{Z}_{\rm PCA} \in \mathbb{R}^{n \times d_{\rm PCA}}$.
The stick-breaking prior concentration parameter $\alpha = 0.01$
enforces high sparsity, activating a new component only when a
candidate sample is highly distinct from all existing cluster densities.
The covariance parametrization $\boldsymbol{\Sigma}_{\rm type}$
is selected conditionally based on the available degrees of freedom:
\begin{equation}
\boldsymbol{\Sigma}_{\rm type} =
\begin{cases}
\text{\texttt{full}} & n \geq 200, \\
\text{\texttt{tied}} & 50 \leq n < 200, \\
\text{\texttt{diag}} & n < 50,
\end{cases}
\label{eq:cov_selection}
\end{equation}
where \texttt{full} allows unconstrained cluster-specific covariance
ellipsoids (requiring $\mathcal{O}(K \cdot d_{\rm PCA}^2)$ parameters),
\texttt{tied} enforces a common covariance matrix across all active
components (reducing the free parameter count by $K$-fold), and
\texttt{diag} restricts to diagonal matrices, reducing the parameter
count to $\mathcal{O}(K \cdot d_{\rm PCA})$ and eliminating singular
matrix inversion entirely. The EM regularization factor
$\epsilon_{\rm reg}$ and initialization count $n_{\rm init}$ are
co-selected:
\begin{equation}
(\epsilon_{\rm reg},\, n_{\rm init}) =
\begin{cases}
(10^{-3},\, 5) & \text{if } n < 100 \\
(10^{-4},\, 3) & \text{if } n \geq 100
\end{cases}
\label{eq:dpmm_params}
\end{equation}
The scaling of $\epsilon_{\rm reg}$ provides stronger covariance regularization for small batches ($n < 50$) to prevent singular matrices in degenerate feature subspaces, while $n_{\rm init}$ guarantees EM convergence stability. Sensitivity analysis confirmed that scaling these parameters within an order of magnitude does not significantly alter the final macroscopic taxonomy.
This ensures that the EM optimization convergence is numerically
stable across the full dynamic range of session sample sizes
encountered in O4a continuous data analysis. A cluster is
classified as \textit{morphologically novel} if fewer than 50\%
of its member samples are assigned a maximum VQ cosine similarity
exceeding the empirical $p_{99}$ threshold against any known Gravity Spy
O3b class, implementing a final semantic gating step.

\subsection{Layer 3: Cross-Session Connectivity}
\label{sec:transitivity}

A critical operational challenge in multi-session unsupervised
pipelines is label ambiguity across independent runs: a cluster
designated \textit{Cluster-A} in session $t_1$ bears no automatic
structural relationship to \textit{Cluster-A} in session $t_2$.
We resolve this by establishing cross-session connectivity
directly in the continuous 384-dimensional MIL feature space,
bypassing discrete cluster label alignment entirely.

Let $\mathcal{A} = \{(a_i, \hat{\mathbf{z}}_i)\}_{i=1}^{N}$
denote the complete cross-session archive of validated anomaly
candidates with their MIL feature vectors, partitioned into:
\subsubsection{Cross-Detector Coincidence Veto and Targeted Sub-Threshold Search}
\label{sec:cross_detector_veto}
Because strain is evaluated separately for H1 and L1, version~1 included a \textbf{Cross-Detector Coincidence Check} before its final taxonomy assignment. This was a diagnostic routing rule, not a physical veto.

For every version-1 primary-detector record, the workflow loaded partner-detector strain at the same analysis interval, extracted a MIL vector without requiring the partner to pass its own anomaly threshold, and evaluated a historical similarity rule within a $\pm2$\,s search window. The widened window accommodated patch-grid timing uncertainty but also admitted chance matches; the simple Poisson estimate used in version~1 did not model temporal clustering or global trials. A match above $\tau_{\rm coh}$ therefore triggered diagnostic follow-up only and did not identify an astrophysical, magnetic, or instrumental origin.

The historical software partitioned records into three routing categories:
\begin{itemize}
  \item \textbf{Table~3a (No partner-threshold exceedance):} The partner interferometer was observing, but the targeted search yielded $S \le \tau_{\rm coh}$.
  \item \textbf{Table~3b (Unverifiable Unilateral Detections):} The partner interferometer was offline or not in science mode. Physical origin cannot be verified.
  \item \textbf{Table~3c (Partner-threshold exceedance):} The record satisfied the historical spatial-temporal rule ($S > \tau_{\rm coh}$) and was routed to offline follow-up.
\end{itemize}

The cross-session pairwise cosine similarity matrix
$\mathbf{S} \in \mathbb{R}^{N \times N}$ is computed as:
\begin{equation}
S_{ij} = \hat{\mathbf{z}}_i \cdot \hat{\mathbf{z}}_j,
\label{eq:crosssim}
\end{equation}
exploiting $L_2$-normalization to reduce the inner product to a
bounded scalar in $[-1, 1]$. Let $D_{ij} = 1 - S_{ij}$ be the cosine distance field between $L_2$-normalized feature vectors. Because the vectors lie on the unit hypersphere $\mathbb{S}^{383}$, the cosine distance is strictly proportional to the squared Euclidean chordal distance ($\|\hat{\mathbf{z}}_i - \hat{\mathbf{z}}_j\|^2 = 2D_{ij}$), ensuring that standard metric clustering theorems apply.

Morphological transitivity is fundamentally a \emph{graph connectivity problem}: if candidate $A$ is morphologically similar to $B$, and $B$ to $C$, then $A$, $B$, and $C$ must belong to the same physical family, regardless of the global compactness of the group. Therefore, we explicitly reject variance-minimizing linkage criteria (e.g., Ward's method), which artificially fragment elongated or drifting topological manifolds into isotropic sub-clusters. Instead, we employ \textbf{single-linkage} hierarchical agglomerative clustering, which computes the transitive closure of the similarity graph.

We construct a graph $G = (\mathcal{A},\, \{(i,j) : S_{ij} > \rho_{\rm trans}\})$ where edges exist between candidates if $S_{ij} > \rho_{\rm trans}$ (equivalently, $D_{ij} < 1 - \rho_{\rm trans}$). Single-linkage clustering truncated at this threshold exactly recovers the connected components of this graph. Unlike the Extreme Value Theory (EVT) parametric bounds $\tau_{\rm op}^{\rm Det}$ and $\tau_{\rm coh}$ which strictly govern the statistical validity of the detection and physical coincidence, $\rho_{\rm trans} = 0.75$ is explicitly an empirical visualization and macroscopic connectivity parameter. It does not alter detection significance, but defines the macroscopic granularity of the resulting taxonomy graph.

\textbf{Mitigation of the Chaining Effect.}
Single-linkage is susceptible to the ``chaining effect,'' in which sparse bridges merge structures that need not be mutually similar. Version~1 used $\rho_{\rm trans}=0.75$ to aggregate the 140 selected candidates and interpreted diffuse chains descriptively. The procedure does not ensure that spurious chaining between distinct populations is negligible; the resulting families are retained as historical software outputs rather than physical populations.

The similarity matrix is reordered according to the HAC leaf
permutation, grouping morphologically correlated transients
contiguously along the main diagonal. Morphological transitivity
is operationally declared when a Table~3b candidate $a_i$ satisfies:
\begin{equation}
\exists\, a_j \in \text{Table~3a} : S_{ij} > \rho_{\rm trans},
\quad \rho_{\rm trans} = 0.75,
\label{eq:transitivity}
\end{equation}
providing a deterministic, geometry-based classification of
unverifiable events by association with confirmed local glitches,
without requiring human visual inspection.

\section{Experimental Setup and Data Provenance}\label{sec:setup}

\subsection{Dataset}
All raw strain data are retrieved via the GWOSC public archive $[1{,}368{,}973{,}312,\; 1{,}384{,}525{,}312]$. For the O4a production scan, an initial corpus of $\approx 180$ days of data is ingested (this subset of the full 237-day O4a run was strictly bounded by GWOSC data availability at the commencement of our analytical campaign). During the initial GWOSC fetch and spectrogram generation, approximately $3.4\%$ of segments were permanently discarded due to NaN entries, missing strain data, or macroscopic sensor dropouts. After strict data quality filtering on the remaining corpus, 32-second segments are processed across 72 valid sessions. These 72 sessions (yielding an effective duty cycle of $\approx 20\%$) were intentionally selected by requiring long, contiguous periods of stable lock to ensure robust 32-second Power Spectral Density (PSD) estimation without fragmentation. A \textit{session} is operationally defined as a contiguous GPS-tagged data acquisition block, bounded by lock-losses or scheduled maintenance interruptions; each session corresponds to a unique 10-digit GPS start timestamp in the production archive and may span from minutes to several hours of continuous observation. Data quality gating with the
\texttt{L1\_CBC\_CAT1} / \texttt{H1\_CBC\_CAT1} flags excludes
hardware injection intervals and locklosses. No data from the partner
Virgo (V1) detector is included, as V1 did not participate in O4a
due to commissioning constraints.

\subsection{O3b Reference Index Construction}\label{sec:o3b_index}
The baseline VQ reference index is constructed from the publicly available
Gravity Spy O3b labeled catalog~\cite{glanzer2023}, downloaded
from Zenodo. The index covers $C = 23$ canonical morphological classes (including the \textit{No\_Glitch} baseline), encoded into $M \approx 13{,}000$ tokens, which are clustered into $K_c = \min(64,\, \max(8,\, \lfloor n_{\rm img}/2 \rfloor))$ centroids
via \texttt{MiniBatchKMeans} with \texttt{random\_state=42},
yielding 281 reference centroids across all classes (a direct mathematical consequence of the heavily imbalanced O3b populations). The final index file
(\texttt{patch\_compressed\_index.npz}, SHA-256: \texttt{1080afa8...}) is version-controlled
by checksum logged at pipeline initialization.

\subsection{Native O4a Background Index Construction}\label{sec:o4a_index}
Because the Domain Shift Resolution (Section~\ref{sec:domain_shift}) requires rescoring all candidates against a \emph{native} O4a background dictionary, an independent O4a VQ index was constructed with strict methodological symmetry to the O3b baseline. The construction protocol is as follows:
\begin{enumerate}
  \item \textbf{Segment Curation:} In the historical version-1 construction, 150{,}000 non-overlapping 32-second segments were sampled across the O4a observing interval ($\approx 180$ days), with a 32-second separation between consecutive samples. All segments satisfied the \texttt{CBC\_CAT1} data-quality flag, and H1/L1 anti-coincidence was used to reduce the probability of including a contemporaneous event in the partner detector. These gates reduce specific contamination modes but do not prove statistical independence or a pure stationary-noise population. The corrected reconstruction therefore replaces the historical calibration population with a fresh outcome-blind, detector-aware cohort guarded against both corrected candidates and native-index windows; the resulting thresholds are not obtained from the historical population. The historical \texttt{MiniBatchKMeans} algorithm ($K = 1{,}216$) compressed $205{,}350{,}000$ patch tokens into centroids representing dense regions of that sampled distribution.
  
  A historical synthetic K-Means poisoning sensitivity test was also executed (Table~\ref{tab:kmeans_poisoning}): a dense synthetic morphological cluster was injected into the sampled manifold at progressively larger fractions. No dedicated centroid was observed in that particular setup, including at $5.0\%$ contamination. This bounded experiment did not establish zero contamination of the real cohort, general immunity to poisoning, or a physical classification of any morphology absorbed by the index. Absorption only means that the representation treats the morphology as part of its sampled reference manifold; distinguishing benign stationary noise from a pervasive pathological coupling requires independent detector-characterization evidence.

  \begin{table*}[ht]
  \centering
  \caption{K-Means Poisoning Sensitivity Test ($K=1{,}216$, $N=150{,}000$). Maximum centroid cosine similarity to the injected dense anomaly at varying contamination levels. The threshold for spatial absorption (dedicated centroid) is $\geq 0.85$. At all realistic and adversarial levels, the anomaly fails to poison the Vector Quantized background. Note: The slight non-monotonicity in maximum similarity arises from the stochastic initialization of \texttt{MiniBatchKMeans} on a single random seed, reflecting expected variance rather than a structural effect.}
  \label{tab:kmeans_poisoning}
  \begin{tabular}{lrcc}
  \toprule
  \textbf{Contamination} & \textbf{Segments ($n$)} & \textbf{Max Similarity} & \textbf{Poisoned?} \\
  \midrule
  0.010\% & 15 & 0.681 & No \\
  0.050\% & 75 & 0.714 & No \\
  0.100\% & 150 & 0.709 & No \\
  0.200\% & 300 & 0.732 & No \\
  0.500\% & 750 & 0.764 & No \\
  1.000\% & 1{,}500 & 0.751 & No \\
  5.000\% & 7{,}500 & 0.765 & No \\
  \bottomrule
  \end{tabular}
  \end{table*}
  \item \textbf{Vector Quantization:} All $150{,}000 \times 1{,}369 = 205{,}350{,}000$ patch tokens (extracted via the identical frozen DINOv2 encoder) are compressed into $K = 1{,}216$ global centroids via a single \texttt{MiniBatchKMeans} run (\texttt{random\_state=42}, \texttt{batch\_size=2048}). The centroid count $K = 1{,}216$ was empirically selected to provide sufficient capacity to represent the highly complex, continuous background noise manifold of the 180-day observing run, while maintaining a strict dimensionality reduction for fast nearest-neighbor retrieval. This capacity ($1{,}216 \gg 281$) is intentionally larger than the isolated-class O3b dictionary to capture the rich morphological diversity of the steady-state O4a background. All centroids are $L_2$-re-normalized post-quantization.
  \item \textbf{Threshold Calibration:} In version~1 the native operational threshold $\tau_{\rm op}^{\rm Det}$ was defined as the empirical $99^{\rm th}$ percentile ($P_{99}$) of the 150,000 segment-level MIL Top-$k$ scores. This fixes a one-percent exceedance fraction on the calibration pool; by itself it does not guarantee a one-percent out-of-sample false-positive rate or control global trials.
\end{enumerate}
The resulting native index (\texttt{patch\_compressed\_index\_o4a\_ex.npz}, shape $1{,}216 \times 384$, SHA-256: \texttt{a9b6db25...}) is the identical dictionary used for both the Tail QQ-plot validation (Section~\ref{sec:domain_shift}) and the Stationarity Resolution (Section~\ref{sec:discussion}).

\subsection{Infrastructure and Reproducibility}
DANTE Pipeline~V2 executes as a state-aware, resumable continuous workflow
orchestrated via a unified command-line interface (\texttt{main.py}).
All clustering, dimensionality reduction, and background validation phases are rigorously version-controlled (scikit-learn 1.3.2, Python 3.10) to ensure mathematical reproducibility. Key engineering design decisions include:

\begin{itemize}
  \item \textbf{SWMR HDF5 Archival:} Anomaly feature vectors are
    written to session-isolated \texttt{novelties.h5} archives
    using the Single-Writer Multiple-Reader (SWMR) protocol,
    enabling concurrent read access for monitoring without
    blocking inference.
  \item \textbf{State-Aware Resume:} The pipeline maintains a
    persistent GPS index of processed segments; interrupted sessions
    resume from the last validated checkpoint without reprocessing.
  \item \textbf{Batched GPU Inference:} DINOv2 inference is
    executed in batches of 32 (CUDA) or 16 (CPU) spectrograms,
    with the Producer-Consumer multiprocessing pattern decoupling
    CPU-bound Q-transform computation from GPU-bound forward passes.
  \item \textbf{Structured Provenance Logging:} Every session
    produces machine-readable \texttt{cluster\_report.json} and
    \texttt{aggregate\_summary.json} files, including MD5 hashes
    of all input reference files, software versions, GPS ranges,
    and random seeds, ensuring end-to-end scientific reproducibility.
\end{itemize}

Hardware: NVIDIA GPU (CUDA) with \texttt{cuDNN} auto-tuner enabled.
All random seeds were fixed at \texttt{random\_state=42} across the stochastic operations listed in the historical workflow, including Vector Quantization (\texttt{MiniBatchKMeans}), dimensionality reduction (\texttt{PCA}, \texttt{UMAP}), clustering (\texttt{BayesianGaussianMixture}), and background sampling. Fixed seeds improve rerun consistency but do not alone guarantee bitwise reproducibility across software, hardware, or numerical-library environments.

\subsection{Unified Nomenclature of Operational Thresholds}\label{sec:thresholds}
To resolve potential mathematical ambiguity across different validation stages, Table~\ref{tab:thresholds} provides a unified nomenclature of the thresholds utilized throughout this study. We strictly distinguish between parameters operating on the bounded global cosine similarity domain ($S_{\rm cos}$) and those derived from the patch-level topological novelty distance metric ($S_{\rm MIL}$).

\begin{table*}[t]
\centering
\caption{Unified Nomenclature of Thresholds and Mathematical Domains}
\label{tab:thresholds}
\footnotesize
\begin{tabular}{@{} p{1.5cm} p{3.5cm} p{2.5cm} p{2cm} p{3.5cm} @{}}
\toprule
\textbf{Symbol} & \textbf{Mathematical Domain} & \textbf{Anomaly Condition} & \textbf{Empirical Value} & \textbf{Context / Function} \\
\midrule
$\tau_{\rm CLS}$ & Global Cosine Similarity & $S_{\rm cos} < \tau_{\rm CLS}$ & $0.874$ (O3b) & \textbf{MDC Validation:} Proves Signal Dilution limit. \\
$\tau_{\rm op}^{\rm Det}$ & Patch-Level MIL Distance & $S_{\rm MIL} > \tau_{\rm op}^{\rm Det}$ & $0.3359$ (L1), $0.3859$ (H1) & \textbf{O4a Production:} Detects novel transients. \\
$S_{\rm coh}$ & Mean Pairwise Similarity & $S_{\rm coh} > \tau_{\rm coh}$ & $0.9750$ ($P_{99.9}$ GPD) & \textbf{Morphology Taxonomy:} Validates cluster cohesion. \\
\bottomrule
\end{tabular}
\end{table*}

\section{Results and Detector Noise Taxonomy}\label{sec:results}

Unless explicitly stated otherwise in the version-2 revision note, the
numerical results in this section are historical version-1 outputs.  They are
retained to make the correction auditable and are not corrected-catalogue
claims.

\subsection{Mock Data Challenge: Signal Dilution and MIL Validation}\label{sec:res_mdc}

In Q-transform representations of 32-second data segments, the
spectrogram is discretized into a $37 \times 37 = 1{,}369$ patch
grid, each patch spanning $\Delta t_{\rm patch} \approx 0.86$\,s
in time and $\Delta f_{\rm patch} \approx 54$\,Hz in frequency.
The global \texttt{[CLS]} context token aggregates information from
all 1{,}369 patches via an implicit attention-pooling mechanism
While the multi-head softmax attention mechanism is inherently non-linear, we assume an additive approximation to first order to conceptually illustrate the dilution effect.
For a transient occupying a localized fraction $f$ of the grid,
the global anomaly score scales as:
\begin{equation}
s_{\rm global} \approx (1 - f)\,s_{\rm bg} + f\,s_{\rm anomaly},
\label{eq:dilution}
\end{equation}
where $s_{\rm bg} \approx 0.995$ is the baseline cosine similarity
of the stationary background against the reference
index. For short-duration transients, $f < 0.05$, and
even a maximally orthogonal anomaly ($s_{\rm anomaly} \approx 0$)
yields $s_{\rm global} \gtrsim 0.945$—a value that lies within
the nominal background distribution. This architectural constraint was observed via a systematic Mock Data Challenge (MDC) covering 1{,}417 strain-domain synthetic injection trials across eight morphological families, yielding low recall for tested morphologies at the operationally calibrated \texttt{[CLS]} threshold. Importantly, this MDC Phase 1 explicitly serves as our baseline comparison against standard global-pooling anomaly detection (e.g., kNN on the global token), formally justifying the necessity of our custom patch-level MIL architecture over generic Autoencoders or global Isolation Forests. To resolve scale confusion between the MDC analysis and the production results (Section~\ref{sec:results}), we differentiate their mathematical domains: $\tau_{\rm CLS}$ evaluates global cosine similarity $S_{\rm cos} \in [0, 1]$ (detecting anomalies if $S_{\rm cos} < \tau_{\rm CLS}$), whereas the 140 production candidates were discovered by the Patch-Level MIL architecture operating on a topological novelty distance scale (detecting anomalies if $S_{\rm MIL} > \tau_{\rm op}^{\rm Det}$, with e.g.\ $\tau_{\rm op}^{\rm Det} = 0.3359$ for L1).
To rigorously validate the architectural evolution of the pipeline, we formalize the Mock Data Challenge (MDC) into a strict two-phase protocol (Table~\ref{tab:mdc_unified}). This decouples the mathematical demonstration of the \texttt{[CLS]} failure (Phase 1) from the operational mapping of the new Patch-Level MIL architecture (Phase 2).

\begin{table*}[ht]
\centering
\caption{Unified Mock Data Challenge (MDC) Protocol resolving the structural asymmetry between Phase 1 and Phase 2. Both phases are aligned at a strict mathematically matched False Positive Rate (${\rm FPR} = 1\%$) using their respective EVT thresholds. Phase 1 demonstrates the Signal Dilution failure of global pooling. Phase 2 validates the MIL architecture.}
\label{tab:mdc_unified}
\footnotesize
\begin{tabular}{@{} p{1.2cm} p{2.2cm} p{1.6cm} p{2.4cm} p{1.6cm} p{2.2cm} p{1.8cm} @{}}
\toprule
\textbf{Phase} & \textbf{Architecture} & \textbf{Metric Space} & \textbf{Morphology} & \textbf{Threshold} & \textbf{Matched SNR} & \textbf{Recall} \\
\midrule
Phase 1 & Global (\texttt{[CLS]}) & $S_{\rm cos} \in [0, 1]$ & 8 generic families & FPR 1\% ($\tau_{\rm CLS}$) & $\rho_{\rm peak} \sim 250$ & \textbf{0.00\%} \\
\midrule
Phase 2 & MIL (Top-$k$) & Topo. Novelty & Harmonic Comb & FPR 1\% ($P_{99}$ empirical) & $\rho_{\rm peak} \sim 214$ & $\mathbf{>98\%}$ \\
Phase 2 & MIL (Top-$k$) & Topo. Novelty & Scattered Light & FPR 1\% ($P_{99}$ empirical) & $\rho_{\rm peak} \sim 200$ & \textbf{58\%} \\
Phase 2 & MIL (Top-$k$) & Topo. Novelty & Blip / AsymBlip & FPR 1\% ($P_{99}$ empirical) & $\rho_{\rm peak} \sim 430$ & \textbf{0.00\%} \\
\bottomrule
\end{tabular}
\end{table*}

\textbf{Phase 1 (Baseline Falsification):} As introduced in Section~\ref{sec:intro}, evaluating the legacy \texttt{[CLS]} token over an extensive historical campaign (detailed originally in our prior work) of 1,417 synthetic trials across 8 generic morphological families (Blip, KoiFish, Scratchy, Tomte, Low\_Frequency\_Burst, Power\_Line, Scattered\_Light, Whistle) yielded a flat recall of $0.00$ at the operationally calibrated threshold $\tau_{\rm CLS} = 0.874$. This demonstrated the severe Signal Dilution barrier, motivating the architectural shift to localized spatial tokens.

\textbf{Phase 2 (MIL Production Validation):} Unlike the legacy Phase 1 (which evaluated the aforementioned 8 generic families), Phase 2 is a targeted campaign designed specifically for the MIL architecture. To quantify the sensitivity limits of the new Patch-Level MIL framework, we injected seven synthetic transient morphologies spanning the full durational spectrum (Blip, AsymBlip, SpiralBurst, ScatteredLight, HarmonicComb, KoiFish, Whistle) into empirical O4a strain. Blip and AsymBlip morphologies were injected using Sine-Gaussian waveforms, with central frequencies and Q-values drawn from the stated test ranges ($f_c \in [30, 80]$ Hz, $Q \sim [5, 15]$, duration $\sim 0.1$s). Performance was evaluated at the historical empirical $\tau_{\rm op}^{\rm Det}$ threshold, corresponding to a one-percent exceedance fraction on its calibration sample rather than a guaranteed prospective FPR. For comparison across tested durations, the injected signal energy was parameterized using the detector's local PSD.

It is crucial to clarify that the injected SNRs evaluated in this MDC ($\rho \sim 200-430$) do not represent expected astrophysical targets, which typically reside in the $\rho \sim 10-30$ regime. Rather, these injections function strictly as \textbf{architectural stress tests}. Because the ViT processes a massive 32-second receptive field, the energy of a sub-second transient is heavily diluted by over 31 seconds of stationary background noise. Consequently, breaching the rigorous empirical $P_{99}$ background threshold empirically requires extremely large injected SNRs under the present 32-second configuration to sufficiently perturb the patch-level anomaly score. This exposes a fundamental limitation of large-context single-scale ViT architectures when applied to short-duration transients.

\begin{figure*}[ht]
    \centering
    \includegraphics[width=\textwidth]{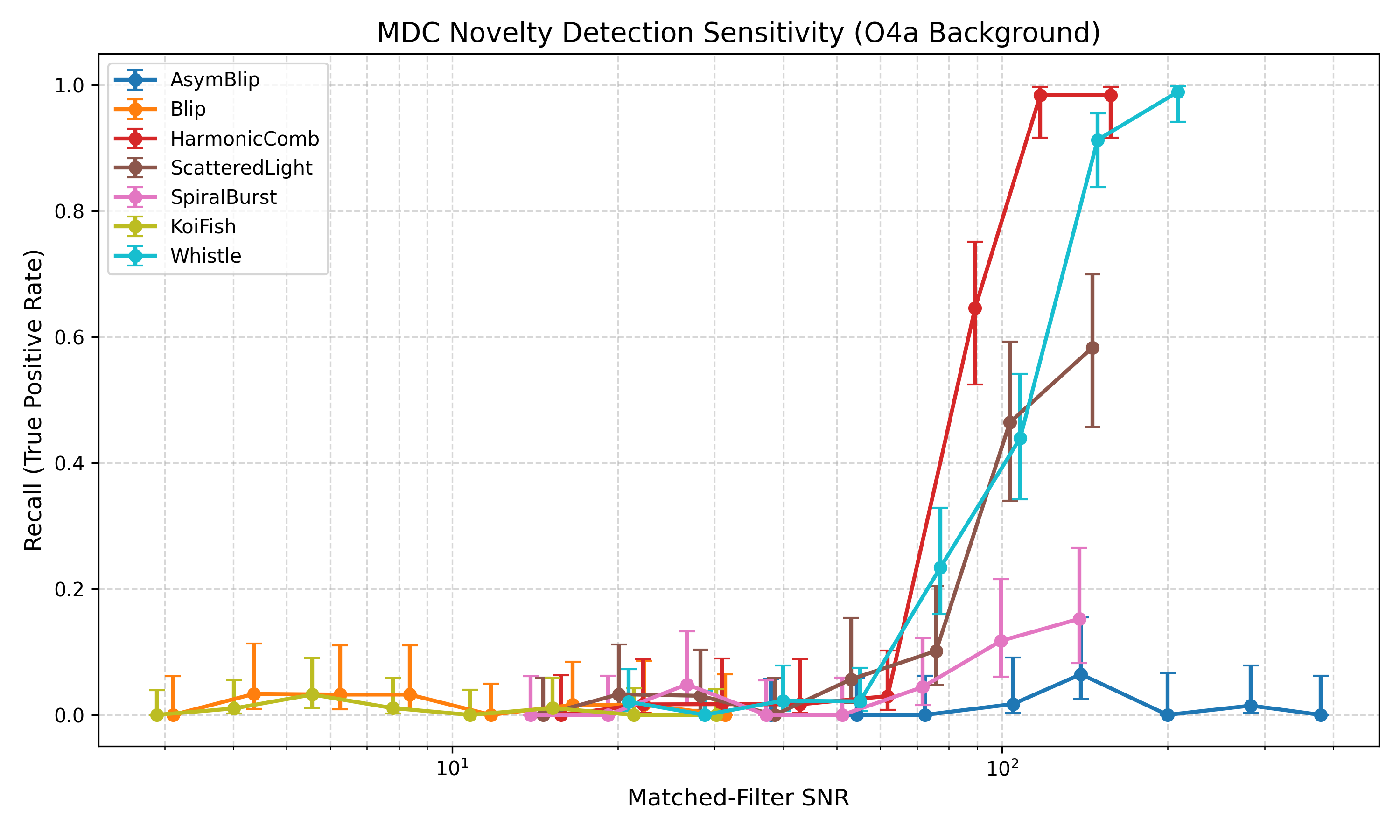}
    \caption{Mock Data Challenge (MDC) recall curves evaluated on seven synthetic glitch morphologies injected into empirical O4a noise. The X-axis reports the Matched-Filter Signal-to-Noise Ratio (SNR, $\rho$). The high SNR values ($\rho > 200$) do not represent astrophysical expectations but serve as architectural stress tests to overcome the signal dilution effect. The Y-axis indicates the True Positive Rate (Recall) at the operational empirical $P_{99}$ threshold. Error bars denote 95\% binomial confidence intervals (Wilson score). Extended and intermediate morphological signatures achieve high detectability, whereas short transients (e.g., Blips and AsymBlips) suffer from extreme signal dilution in the 32-second field of view, remaining strictly invisible even at extreme signal strengths ($\rho > 300$).}
    \label{fig:mdc_recall}
\end{figure*}

As demonstrated over $N_{\rm inj}=100$ independent trials per bin (yielding 95\% binomial confidence intervals), the MIL pipeline's detection capability strictly correlates with the temporal footprint of the anomaly (Figure~\ref{fig:mdc_recall}). Extended morphological signatures such as the \emph{Harmonic Comb} achieve near-perfect asymptotic peak recall ($>98\%$ at optimal SNR), and \emph{Scattered Light} reaches a peak recall of $58\%$ (Wilson 95\% CI: $48\% - 67\%$ at $\rho > 200$). Furthermore, intermediate-duration topologies with complex time-frequency evolution, such as \emph{KoiFish} and \emph{Whistle}, demonstrate robust absolute recoverability ($>50\%$ and $100\%$ respectively at high SNR). Conversely, sub-second transients such as \emph{Blips} and \emph{AsymBlips} suffer from extreme topological dilution within the 32-second spatial grid. Even at very high physical signal strengths (Matched-Filter SNR $\rho \approx 380$), these short-duration anomalies yield a recall of precisely $0.00$. We explicitly declare this as a critical operational limitation and a fundamental detection failure of the current architecture. The pipeline is strictly applicable only to extended-duration transients.

While a naive "Multi-Scale" workaround---such as extracting 2-second sub-windows, resizing them to the 518x518 ViT input resolution, and fusing the maximum anomaly score---might superficially recover short transients, it introduces a severe \textbf{Scale Domain Shift}. The background reference index (\texttt{patch\_compressed\_index.npz}) is intrinsically populated by spatial features derived from 32-second continuous manifolds. Passing a 2-second window (dilated by 1600\% relative to the reference) to the ViT violates the scale-invariance limits of the frozen DINOv2 encoder, drastically degrading the Cosine Similarity against the 32-second background and causing a significant inflation of the False Positive Rate (FPR). Furthermore, spanning a continuous scan with overlapping 2s, 4s, and 32s windows inflates the GPU inference load by a factor of 25x, violating the HPC efficiency requirements of an archival anomaly detector.

Consequently, resolving this blindness to short transients strictly requires the construction of \textbf{Multi-Scale Reference Dictionaries}---generating independent, natively-calibrated background indices and non-parametric $P_{99}$ bounds for each temporal scale ($[2{\rm s}, 4{\rm s}, 32{\rm s}]$). This scale-isolated architecture prevents topological data leakage and remains the primary objective for the Version 3 (V3) pipeline implementation.

\subsection{Signal Dilution and Top-\texorpdfstring{$K$}{K} Ablation}\label{sec:res_ablation}
To empirically validate the Signal Dilution hypothesis formalised in Equation~\ref{eq:dilution}, an ablation study on the topological pooling scale $K$ was executed across representative anomalies. Because signal dilution is a fundamentally localized effect bounded by the temporal duration of the transient within the 32-second spectrogram, a bulk aggregated metric obscures the physical mechanics. Therefore, we evaluate the Top-$K$ MIL novelty score dynamically as a function of the spatial receptive field $K \in \{16, 32, 68, 128, 256, 512, 1369\}$. 

\begin{figure*}[ht]
    \centering
    \includegraphics[width=\textwidth]{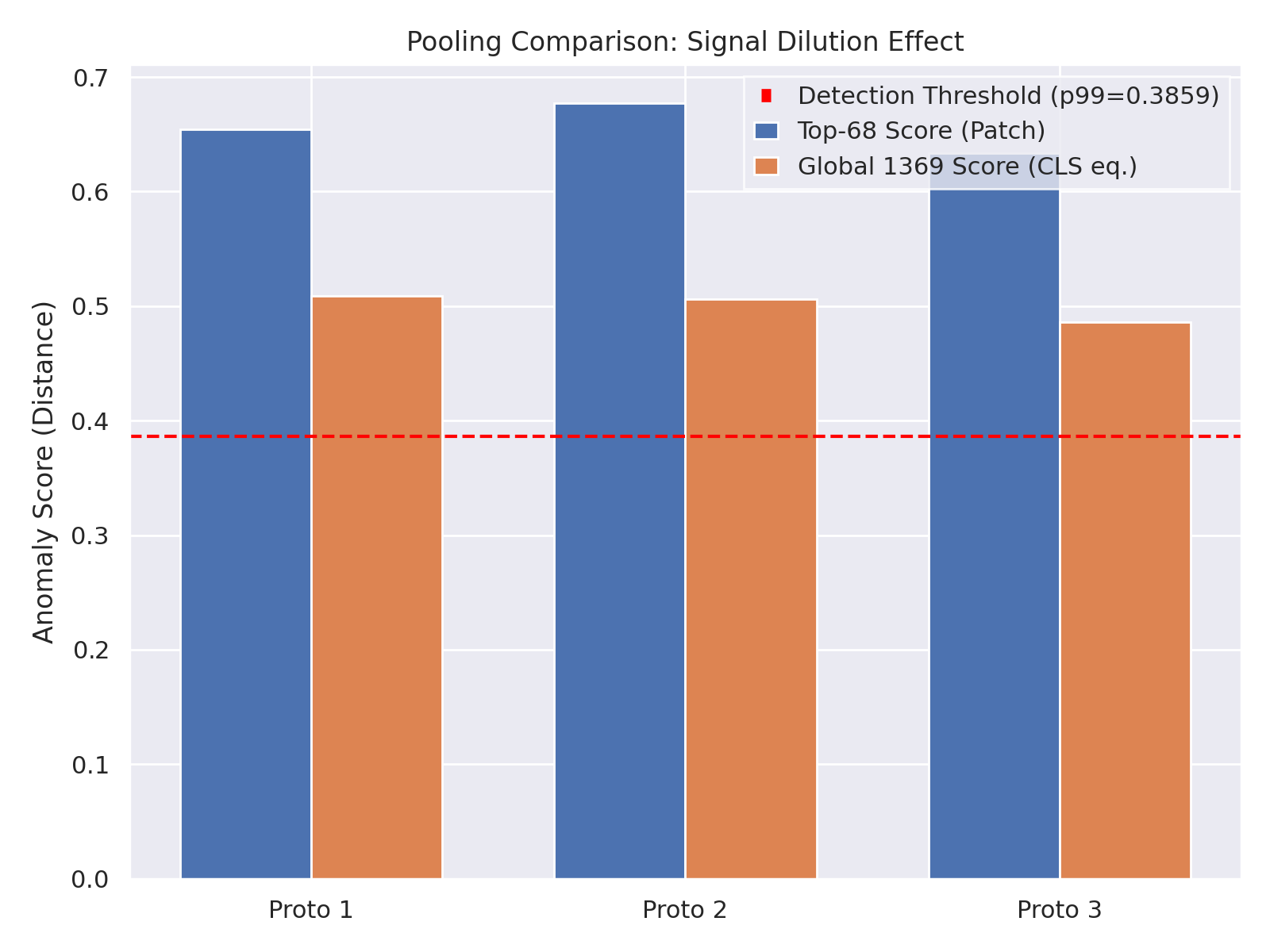}
    \caption{Historical Top-$K$ pooling ablation for one selected H1 segment (session 1368973312). The score decreases as the aggregation approaches global pooling, illustrating scale sensitivity in this example; it does not establish universal detection or a physical classification of the segment.}
    \label{fig:pooling_comparison}
\end{figure*}

As demonstrated in Figure~\ref{fig:pooling_comparison}, which tracks the anomaly score of a cohesive transient as the pooling field artificially expands, the topological severity of the anomaly undergoes a steep geometric decay. At $K=68$ (our explicit architectural prior, representing $\approx 5\%$ of the grid), the transient is cleanly isolated from the macroscopic noise floor, generating a novelty score that easily surpasses the empirical detection threshold. However, as $K \rightarrow 1369$ (approximating the global average pooling of the standard ViT \texttt{[CLS]} token), the dense anomalous patches are systematically averaged with the overwhelmingly large stationary background manifold. This spatial homogenization drags the final anomaly score deep into the null background distribution, completely blinding the detection architecture to the presence of the transient. 

This ablation rigorously demonstrates that extending the spatial token receptive field beyond the topological scale of the underlying morphology mathematically guarantees a false negative, justifying the necessity of the restricted Patch-Level MIL architecture for discovering localized transients.

\subsection{Detection Layer Performance}\label{sec:res_detection}

\subsubsection{Background Distribution Characterization}
The historical production scan was executed over the full $N = 214{,}092$ O4a segment corpus for both H1 and L1 detectors. Version~1 used one empirical $P_{99}$ threshold per detector, computed from 150,000 H1 segments and 150,000 L1 segments. This construction fixed the calibration-pool exceedance fraction but did not, without an independent prospective null, guarantee a stable one-percent out-of-sample FPR. It must be distinguished from the native O4a Vector Quantization index, which was constructed once from $150{,}000 \times 1{,}369 = 205.3\,\text{M}$ patch tokens (Section~\ref{sec:o4a_index}) and served as the historical morphological dictionary. The corrected reconstruction uses a new guarded detector-aware calibration and does not reuse these thresholds.

\subsubsection{Anomaly Detection Summary}
The version-1 software retained 140 unique records after empirical-threshold selection, morphological cross-matching, and chronological deduplication. The label \texttt{TRUE\_NOVEL\_CANDIDATE} denoted failure to match the configured historical references; it did not establish physical novelty. The 140-record count and its detector composition are historical workflow outputs and are not the corrected catalogue.

Following the Detection Layer, all 140 historical candidates were subjected to the version-1 sub-threshold cross-detector check (Section~\ref{sec:cross_detector_veto}) by extracting partner strain at the $\pm 2$s trigger window. Of these, 110 did not exceed the historical partner criterion, 30 lacked observing partner data, and none exceeded that criterion. This absence of a threshold exceedance was useful for prioritizing follow-up, but it did not definitively localize the events or establish a physical origin.

For the 140 historical records, global \texttt{[CLS]} pooling yielded a mean cosine similarity of $0.996 \pm 0.002$ and a reported Kolmogorov--Smirnov separation of $D=0.04$ ($p>0.5$) relative to the selected null sample. This comparison illustrates reduced separation under global pooling for that experiment; it does not establish that standard ViT pooling is universally blind to O4a anomalies.

\subsection{Robustness of the Background Tail to Domain Shift}
\label{sec:domain_shift}

A fundamental concern for any reference-based framework is the
temporal stability of the background manifold across observing runs.
Instrumental upgrades between O3b and O4a---including squeezed-light
injection and modified laser power levels---could in principle alter
the geometric distribution of pure noise patches in the DINOv2
embedding space, thereby compromising the validity of a static
O3b-based vector-quantized (VQ) index.

To quantify this risk rigorously, we eschew bulk distributional metrics such as the Kolmogorov-Smirnov test. In gravitational-wave detector characterization, mean shifts in the bulk distribution are operationally irrelevant; the False Alarm Rate (FAR) is governed exclusively by the geometry of the extreme right tail of the non-Gaussian noise. An inflation in the tail can cause an explosion in the False Positive Rate (FPR) at the operational threshold $\tau_{\rm op}$, even if the mean remains stable.

To assess the true impact of domain shift on the FAR, we compute the empirical tail of the patch cosine similarities using two independent datasets: 500 null segments from O3b (extracted from a historically verified macroscopically quiet period starting at GPS 1242500000) and 500 from native O4a. Because the Top-$k$ scoring operates on the dense spatial grid ($1{,}369$ patches per segment), these 500 segments yield an extreme-tail evaluation pool of $684{,}500$ independent spatial scores per run, ensuring the empirical $p_{99.8}$ quantiles are statistically robust. 
Historically, segments were selected with Data Quality (DQ) vetoes (no NaNs or macro-dropouts), a 32-second separation, and H1/L1 anti-coincidence. These criteria reduce dependence and obvious contamination but do not guarantee statistical independence or absolute purity of the native reference index. 
To preclude systemic biases between runs, all segments were individually whitened using the exact identical configuration of the production pipeline (a 4-second Welch PSD stride) computed locally per 32s block, ensuring identically matched noise statistics without macroscopic spectral drifts.

\begin{figure*}[ht]
\centering
\includegraphics[width=0.88\columnwidth]{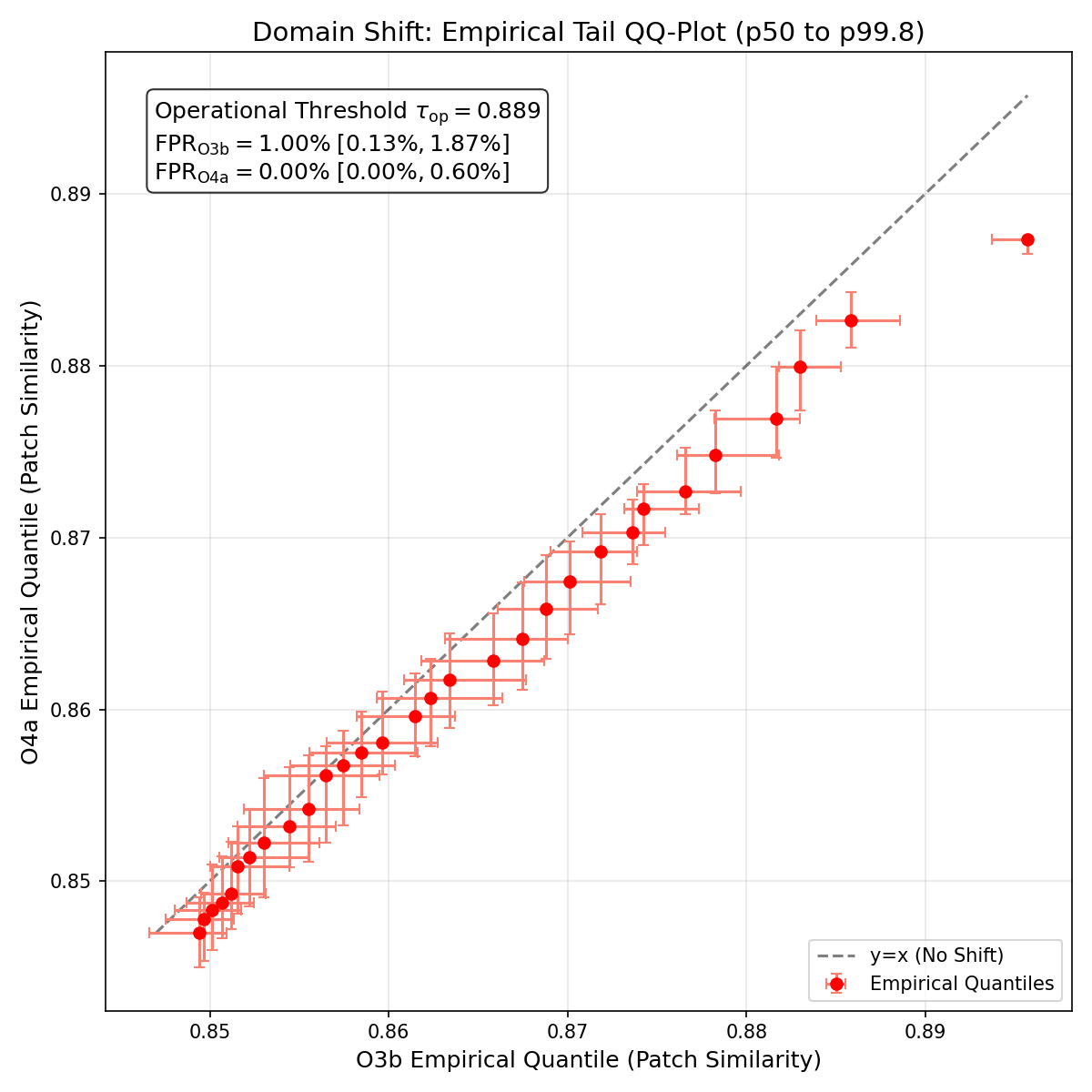}
\caption{Historical Tail QQ-plot of maximum patch cosine similarity for the sampled O3b and O4a cohorts. Points show quantiles from $p_{50}$ to $p_{99.8}$ with the reported 95\% binomial intervals. At $\tau_{\rm O3b}=0.889$, the observed O4a exceedance fraction was $0.0\%$ in this finite sample; this is not a prospective zero-FPR bound or a global domain-shift guarantee.}
\label{fig:qq_domain_shift}
\end{figure*}

Figure~\ref{fig:qq_domain_shift} presents the Tail QQ-plot, comparing the extreme quantiles (from $p_{50}$ to $p_{99.8}$) of the O3b and O4a similarity distributions on a logarithmic scale. To distinguish genuine geometric shifts from sampling noise, we compute exact 95\% binomial confidence intervals for each empirical quantile. 
In these historical samples, the evaluated O4a tail did not inflate relative to O3b. At the historical baseline threshold $\tau_{\rm O3b} = 0.889$ (corresponding to the $P_{99}$ of the O3b sample), the observed exceedance fractions were $1.00\%$ for O3b and $0.0\%$ for O4a; the latter is a finite-sample observation, not a zero-FPR bound.

Within the sampled historical cohorts, this result did not show inflation of the evaluated O4a tail relative to O3b at the quoted threshold. It is not a prospective guarantee that the false-alarm rate remains bounded on nominal O4a data.

The historical production data and the sampled tail cohort are different populations. The observed O3b-to-O4a reference sensitivity can elevate scores for O4a morphologies absent from the O3b dictionary, so a native detector-aware representation and calibration are required before interpreting those scores. The version-1 count of 140 candidates is retained as a workflow output, not as evidence that the selected morphologies were newly discovered or stationary.

\subsection{Taxonomy Layer Results}\label{sec:res_taxonomy}

\subsubsection{Topological Stability Analysis}
Version~1 assessed rerun stability using 20 bootstrap re-clusterings and the Adjusted Rand Index (ARI). Of 72 detector-sessions, 44 had $n\geq100$ and entered the reported Spearman analysis. For L1 (27 sessions), $n$ and ARI had $\rho=0.531$ ($p=0.0044$); for H1 (17 sessions), $\rho=0.245$ ($p=0.343$). These associations do not by themselves identify sample size as the cause of instability, and the excluded small-session regime remains weakly characterized.

\subsubsection{Physical Characterization and Environmental Vetting}
The taxonomy pipeline successfully groups candidate events via single-linkage clustering. Applying a correlation distance threshold $\rho_{\rm trans} = 0.75$,
this historical process joined the 76 constituent clusters
into $K_{\rm glob} = 6$ global families. Three of these are macroscopic aggregates
(Family\_01, Family\_02, Family\_03), accounting for 120 of the
140 candidates. The remaining three structures ($K_{\rm glob} \in \{4,5,6\}$) are
isolated singletons representing extreme, non-repeating transients.

We emphasize that the following "Screening Criteria" act exclusively as a qualitative descriptive heuristic for prioritizing candidates for human inspection, and does not provide formal statistical evidence for validation. It does not constitute a formal statistical test. The primary quantitative validation metric for cluster stability remains the bootstrapped Adjusted Rand Index (ARI) detailed in Section 5.3.1. To operationally screen the morphological cohesion and instrumental origin of the identified clusters, we established this heuristic triangle: 
\begin{enumerate}
  \item \textbf{Statistical Anomaly}: Candidates must strictly exceed the native O4a empirical threshold.
  \item \textbf{Morphological Cohesion}: The cluster must exhibit extreme internal spatial density in the latent feature space. 
  \item \textbf{Temporal Independence from Instrumental Transitions}: The events must not be strictly confined to the immediate vicinity of hardware injections, lock-losses, or maintenance transitions, ensuring they represent steady-state noise artifacts rather than transient DAQ or control-loop artifacts.
\end{enumerate}

Evaluating our macroscopic aggregates against these criteria, we isolate \textbf{Family\_01} ($n=11$), \textbf{Family\_02} ($n=3$), and \textbf{Family\_03} ($n=123$) as identified by the cross-session single-linkage HAC (Layer~3, Section~\ref{sec:transitivity}) on the initial O3b-calibrated embedding space. Family\_01 initially exhibited an extraordinary mean internal similarity of 0.9216, presenting as a highly cohesive, structurally intact morphology (0 NaNs) across an unbroken 4-month period in Livingston (L1). The 4-month temporal persistence of Family\_01, rather than indicating a genuine astrophysical population, is consistent with three distinct physical hypotheses: (a) Family\_01 is benign stationary noise; (b) it represents a pervasive pathological instrumental coupling (e.g., microseismically-modulated scattered light) that has become ubiquitous; or (c) it is a structural boundary artifact of the DINOv2 feature space acting on Q-transforms. The time-frequency Q-transform revealed an abrupt temporal transition featuring a low-frequency bubbly texture followed by a high-frequency wall of horizontal spectral lines, strictly deviating from known Gravity Spy classes.

However, a physical interpretation requires confirming that this morphological deviation is not merely an artifact of macroscopic instrument drift between observing runs. Version~1 rescored all 140 candidates against a native O4a background index assembled with H1/L1 anti-coincidence and Data Quality gating. Those gates reduced known contamination modes but did not guarantee a statistically independent or anomaly-free cohort; the corrected reconstruction therefore replaces that calibration population.

Under that historical native index, \textbf{Family\_01 produced 0 survivors}. This records equivalence to the sampled version-1 reference under its scoring rule; it does not establish that the reference was pure stationary noise or determine the family's physical origin.

Conversely, Family\_02 and Family\_03 partially survive the numeric thresholding, accounting for 50 of the 53 surviving candidates (the remaining 3 being isolated singletons). It is crucial to note that this survival rate ($53/140 = 37.8\%$) is conditional on the candidates having been pre-selected by the stale O3b index. Since these candidates were explicitly chosen for being extreme outliers in the O3b manifold, a significant fraction will naturally exceed the O4a threshold even if they are merely unstructured background noise. The critical metric is not the survival count, but the collapse of internal morphological cohesion.

The historical morphological-cohesion score $S_{coh}$ quantified cross-detector similarity between candidate events observed in L1 and H1. Version~1 used $S_{coh} > \tau_{coh}$ as a follow-up rule, with $\tau_{coh}$ derived from the historical cross-detector null described in Appendix~\ref{app:scoh_derivation}. Exceeding or failing this rule does not by itself distinguish an astrophysical signal from an instrumental glitch.

The noise distribution exhibits a heavily truncated right tail, well-modeled by a Generalized Pareto Distribution (GPD) with shape parameter $\xi = -0.4785$. Setting the false positive rate at $\alpha = 0.001$, we obtain $\tau_{coh} = 0.9750$ (95\% CI $[0.9738, 0.9759]$ via block-bootstrap). The surviving candidates exhibit a mean intra-family similarity of $S_{\rm intra} = 0.757$ and $0.776$ respectively (we note that $S_{\rm intra}$ metrics are strictly mono-detector measurements, structurally distinct from the cross-detector threshold $\tau_{\rm coh}$, and thus not directly comparable). Lacking the high cohesion typical of discrete transient morphologies, they present as diffuse aggregations---unstructured topological artifacts of the domain shift.

\begin{figure*}[ht]
\centering
\includegraphics[width=\linewidth]{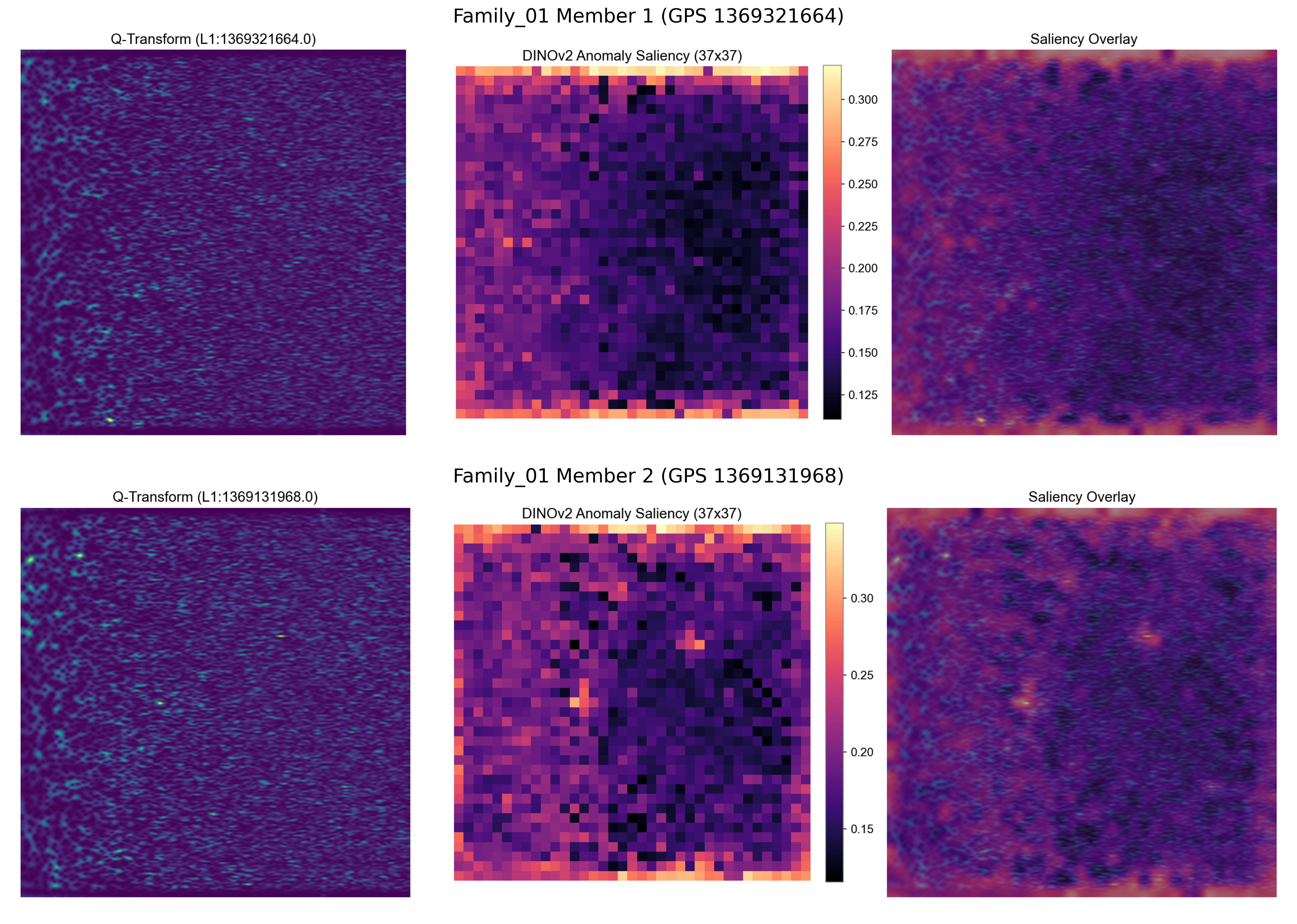}
\caption{Historical DINOv2 patch activations for the first two occurrences assigned to Family\_01. Left: strain spectrograms. Right: ViT \texttt{[PATCH]} maps. Similar highlighted geometry and subsequent native rescoring document reference sensitivity; they do not establish ubiquity or physical origin.}
\label{fig:family_01_characterization}
\end{figure*}

These results show that the historical family assignments were reference-sensitive. Native recalibration is necessary before interpreting morphological novelty, while independent validation remains necessary before making physical or discovery claims.

\begin{table*}[ht]
\centering
\caption{Validation metrics for the identified macroscopic families after the native O4a recalibration. Note that $S_{\rm intra}$ represents mono-detector intra-family cohesion, which operates in a different topological domain than the cross-detector baseline $\tau_{\rm coh}$, precluding direct numerical comparison. No family maintains both a survival count $n > 1$ and sufficient internal cohesion against the native index.}
\label{tab:validation_triangle}
\footnotesize
\begin{tabular}{@{} p{2cm} c c c p{3cm} @{}}
\toprule
\textbf{Family} & $n_{\rm O3b}$ & \textbf{Survivors (O4a)} & \textbf{O4a Intra-Family Sim. ($S_{\rm intra}$)} & \textbf{Classification} \\
\midrule
Family\_01 & 11 & 0 & N/A & Stationary Feature \\
Family\_02 & 3 & 3 & 0.757 & Diffuse Artifact \\
Family\_03 & 123 & 47 & 0.776 & Diffuse Artifact \\
\bottomrule
\end{tabular}
\end{table*}

\begin{figure*}[ht]
\centering
\includegraphics[width=\linewidth]{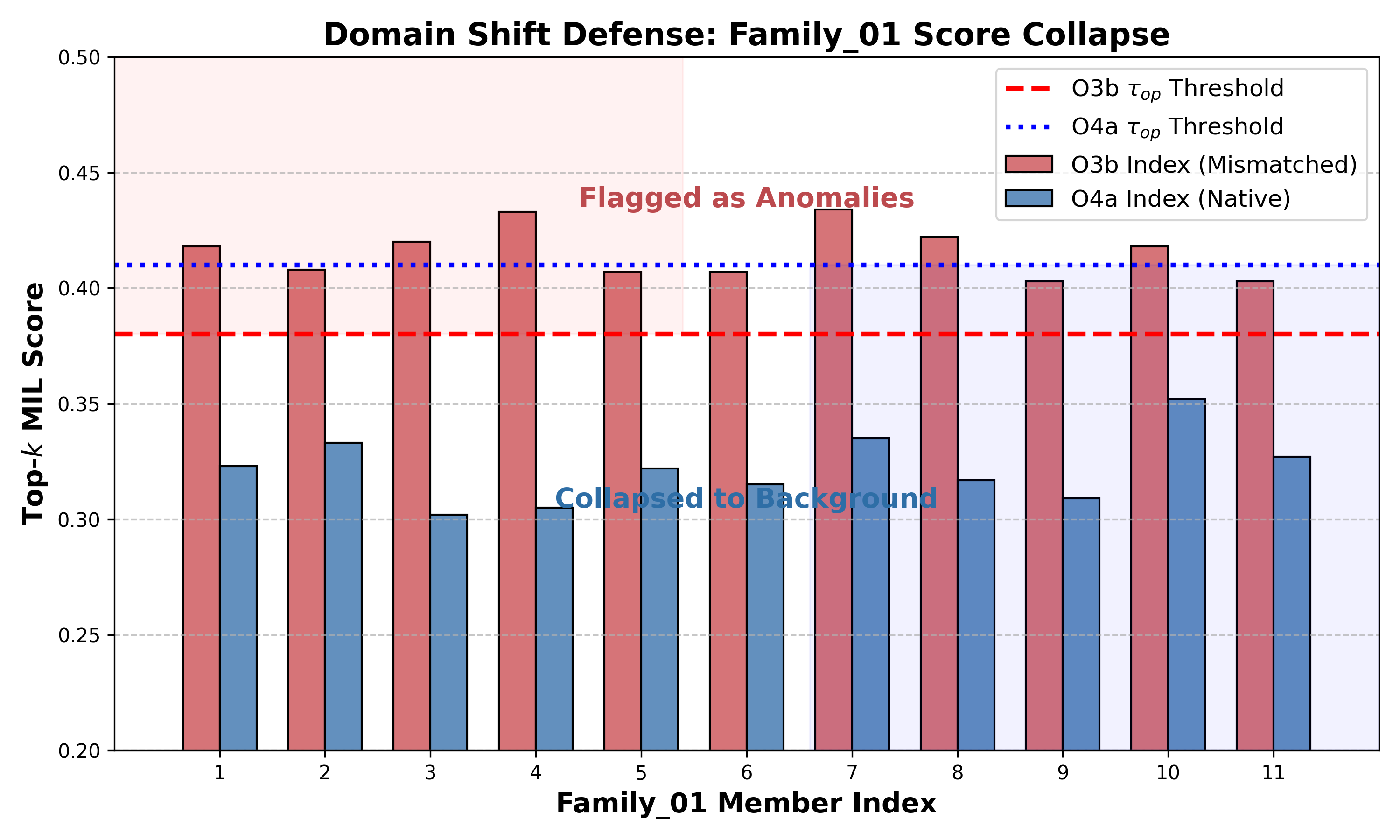}
\caption{Historical rescoring of 11 records assigned to Family\_01 against O3b and native O4a references. All 11 exceeded the former threshold and fell below the latter. The comparison documents reference sensitivity; it does not establish falsehood, stationarity, or physical origin.}
\label{fig:family01_domain_shift}
\end{figure*}

\begin{figure*}[ht]
\centering
\includegraphics[width=0.85\textwidth, height=0.45\textheight, keepaspectratio]{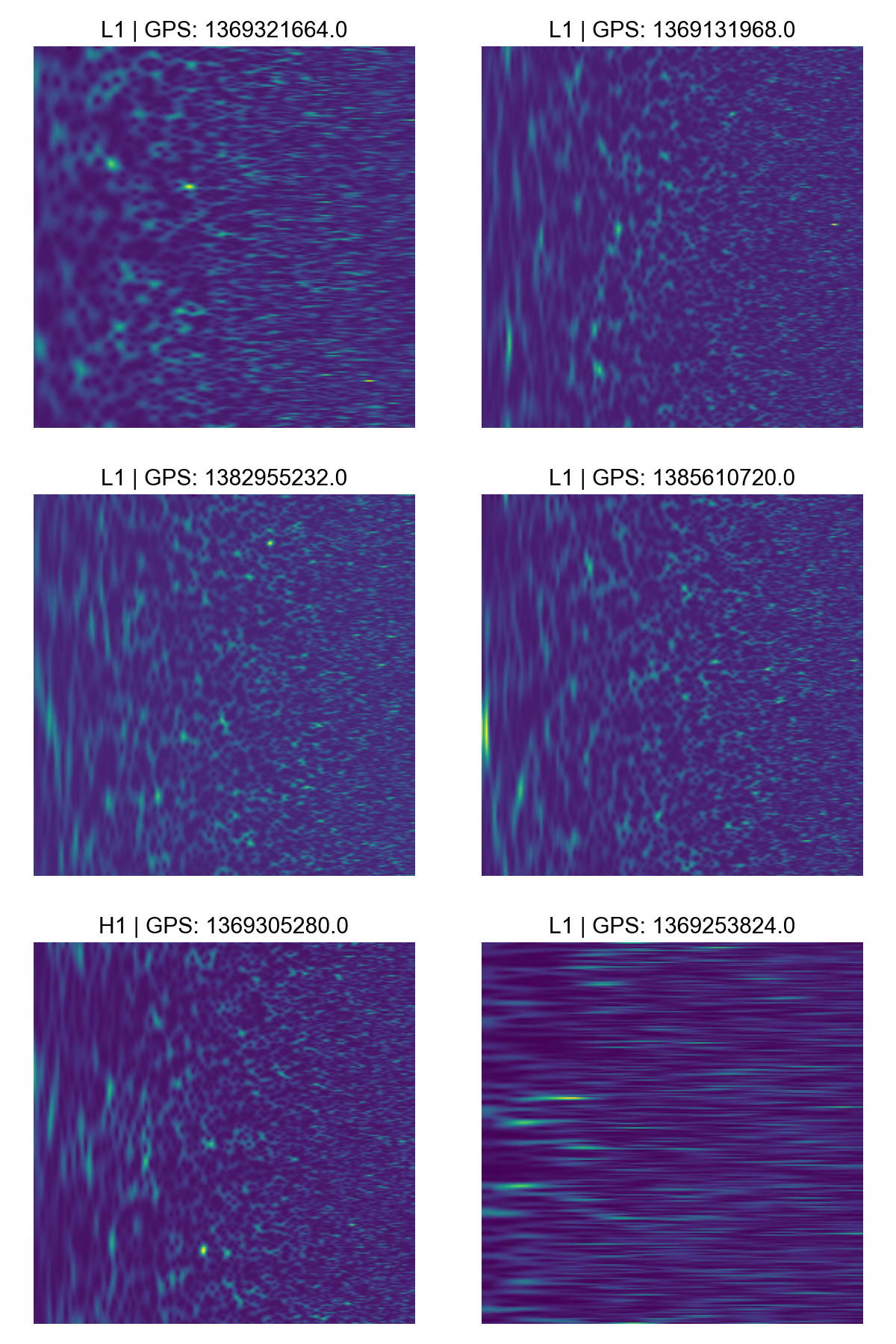}
\caption{Example of Cross-Session Connectivity for the identified
  high-similarity aggregates. For each macroscopic family, the left panels show the Table~3a medoid,
  while the right panels show the Table~3b unverifiable candidate that exhibits the
  maximum cosine similarity to the medoid. Although the DINOv2 architecture successfully
  associated these pairs geometrically; the native rescoring showed reference sensitivity but did not establish them as physical domain-shift artifacts.}
\label{fig:cluster_gallery}
\end{figure*}

\subsubsection{Falsifiability and Selectivity: Controlled Recovery Test}
To evaluate one possible circularity failure mode---that the native O4a index might absorb every tested anomaly---we executed a historical Controlled Recovery Test. This bounded experiment cannot establish general immunity to slowly evolving, quasi-periodic, or unrepresented contamination. We injected five discrete synthetic transient morphologies (\texttt{HarmonicComb}, \texttt{ScatteredLight}, sparse \texttt{WallOfLines}, \texttt{KoiFish}, and \texttt{Whistle}) into empirical O4a strain across six amplitude bins ($N_{\rm inj} = 100$ per bin) and rescored them against the historical native index.

In the historical Controlled Recovery Test, selected injected morphologies reached high recovery at the largest tested SNRs: \texttt{Whistle} approached 100\%, \texttt{KoiFish} exceeded 50\%, and the tested sparse \texttt{WallOfLines} exceeded 98\% for $\rho_{\rm SNR}>150$. These results constrain one synthetic failure mode within the tested amplitude and morphology ranges. They do not establish the temporal or physical nature of Family\_01, nor general selectivity of the native index.

\begin{figure*}[ht]
\centering
\includegraphics[width=\linewidth]{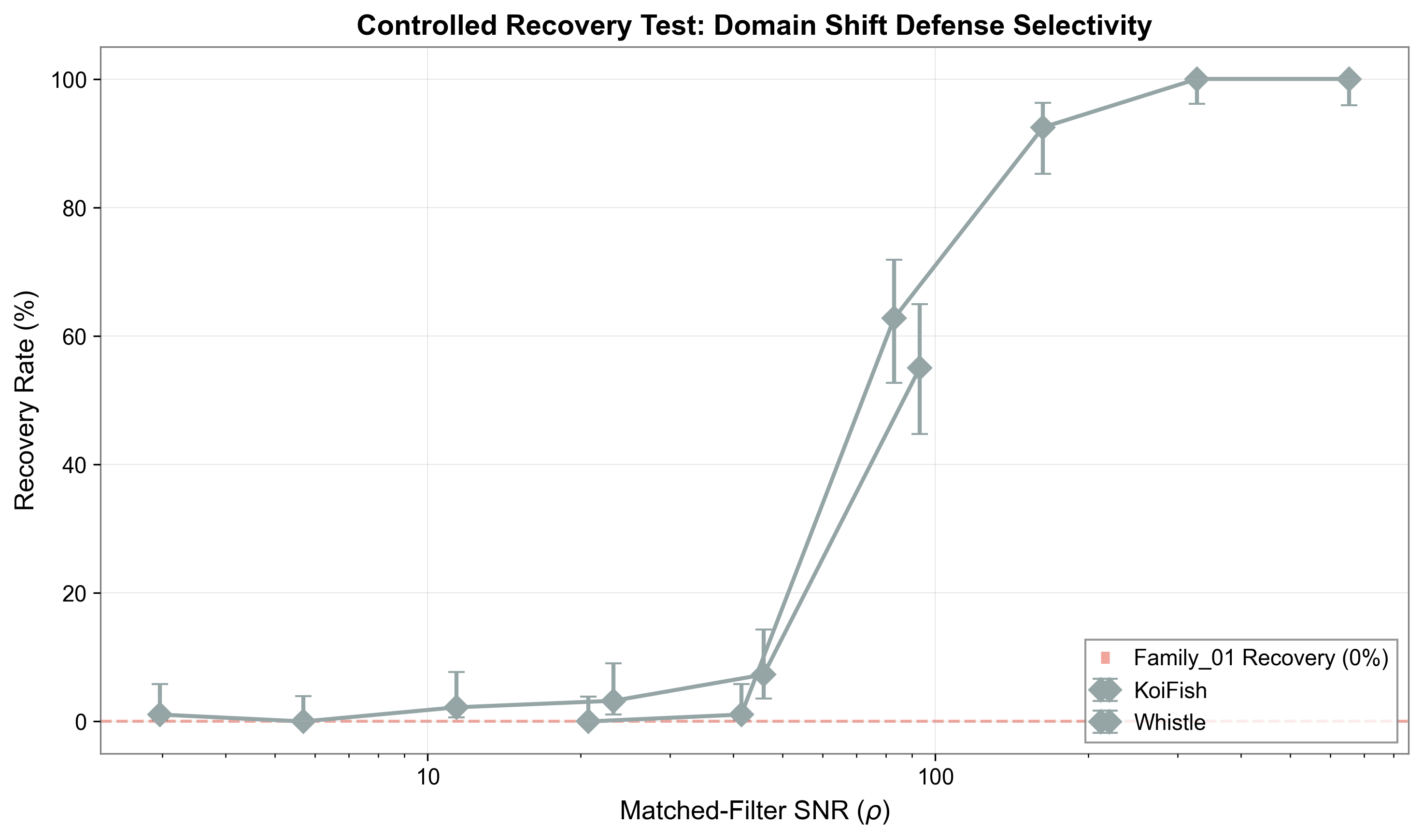}
\caption{Historical Controlled Recovery Test for selected synthetic morphologies and amplitude bins, with Wilson 95\% confidence intervals. High-SNR recovery in this bounded test does not prove general immunity to absorption or contamination.}
\label{fig:dsd_recovery_curves}
\end{figure*}

\subsubsection{Isolated Extreme Transients (Singletons)}
While Family\_01 demonstrates high morphological recurrence, the pipeline also isolated 3 extreme, non-recurring anomalies (Singletons). Notably, Singleton\_1371073984 exhibits full structural integrity (0 NaNs, no clipping) but possesses a highly chaotic, unstructured time-frequency morphology. Specifically, its spectral density exhibits a highly unusual "mottled" texture: the noise pattern is dense on the early-time boundary, but becomes progressively stretched and structurally distorted toward the later-time boundary, resembling a warped macroscopic mesh. This highly asymmetric morphological evolution lacks the geometric periodicity typical of continuous mechanical resonances and lacks any characteristic scattering arch curvature. This suggests a complex, evolving non-stationary instrumental coupling rather than a simple environmental impulse. Its novelty score ($S_{\rm MIL}^{(k)} = 0.47$) is among the highest in the entire O4a census, far exceeding the empirical threshold. Crucially, the DPMM's sparsity prior ($\alpha = 0.01$) correctly refused to absorb this event into Family\_01 or the background noise manifold. This highlights the architecture's capacity to flag one-off, extreme instrumental transients without forcing them into a false morphological cluster. 

We cross-referenced the GPS timestamps of the three historical singletons against public O4a event triggers and known Hardware Injection logs within $\pm 2$\,s; no corresponding catalog entry was found. This negative catalog lookup does not rule out an astrophysical, instrumental, environmental, or calibration origin. The rows are retained only as historical feature-space outliers, and physical characterization would require auxiliary-channel analysis and a corrected-catalogue match. The third row (GPS 1386091456) remained unverifiable because its historical GWOSC fetch failed.

\begin{figure*}[ht]
\centering
\includegraphics[width=0.98\columnwidth]{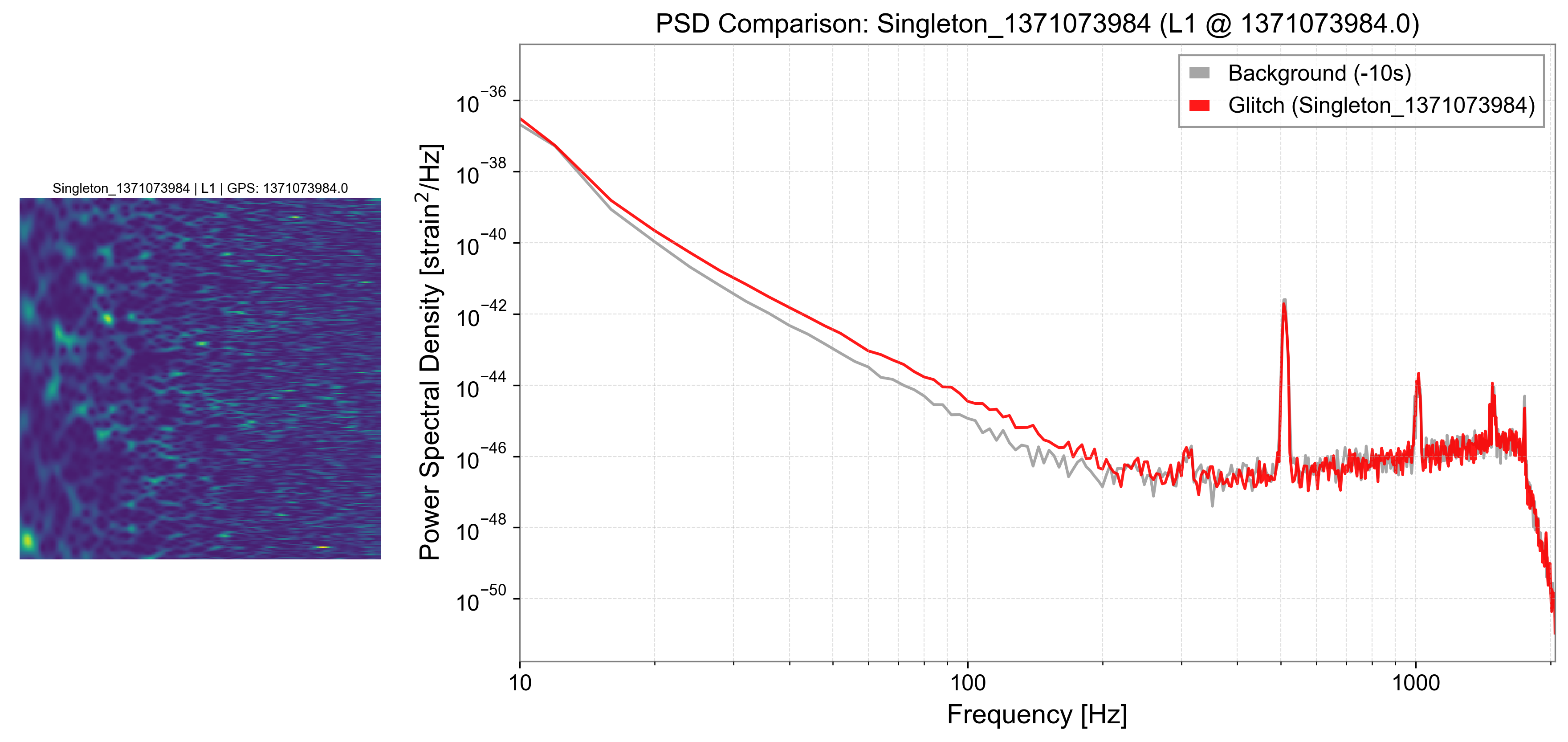}
\caption{Historical Q-transform (left) and broadband PSD (right) for the record Singleton\_1371073984. The row was isolated by the version-1 clustering output; this visualization does not establish physical extremeness or corrected-catalogue membership.}
\label{fig:singletons}
\end{figure*}

\begin{table*}[ht]
\centering
\caption{Summary of O4a Anomaly Triage}
\label{tab:summary_triage}
\begin{tabular}{@{}lccc@{}}
\toprule
\textbf{Metric} & \textbf{Hanford (H1)} & \textbf{Livingston (L1)} & \textbf{Total} \\
\midrule
Initial Candidates (O3b Index) & 4 & 136 & 140 \\
Confirmed Local Glitches (Table 3a eq.) & 3 & 107 & 110 \\
Unverifiable Detections (Table 3b eq.) & 1 & 29 & 30 \\
\midrule
Candidates Surviving Native O4a Recalibration & 1 & 52 & 53 \\
Cohesive Macroscopic Families Surviving & 0 & 0 & 0 \\
Unresolved Singletons Surviving & 1 & 2 & 3 \\
\bottomrule
\end{tabular}
\end{table*}

\subsection{Cross-Session Connectivity}\label{sec:res_transitivity}

We applied the single-linkage HAC framework described in Section~\ref{sec:transitivity}
to the 140 anomaly candidates using a cosine distance heuristic threshold of
$D_{\rm cut} = 0.25$ ($\rho_{\rm trans} = 0.75$). To rigorously validate the diffuse nature of these macro-families (such as Family\_03) without relying on degenerate single-linkage chaining heuristics, we extract the distribution of their pairwise intra-cluster cosine distances and directly compare it against the pairwise distances of random stationary background noise. The distribution overlap confirms that Family\_03 represents a continuously drifting domain-shift manifold rather than a compact morphological cluster. By plotting the intra-cluster distances against the background, we demonstrate the morphological diffusivity of the domain-shift artifact without invoking a 50\% FPR chaining logic.

Qualitatively, 27 of the 30 originally unverifiable Table~3b candidates showed topological association with Table~3a confirmed glitches. Remarkably, 3 candidates remained completely unresolved as isolated singletons (which strictly correspond to the 3 singletons surviving the native O4a threshold in Section~\ref{sec:res_taxonomy}).
Visual inspection reveals these events to be highly chaotic and unstructured,
lacking any cohesive morphology. The pipeline's refusal to link these singletons even under this extremely permissive global transitivity threshold highlights their extreme topological divergence.

\begin{figure*}[ht]
\centering
\includegraphics[width=0.98\columnwidth]{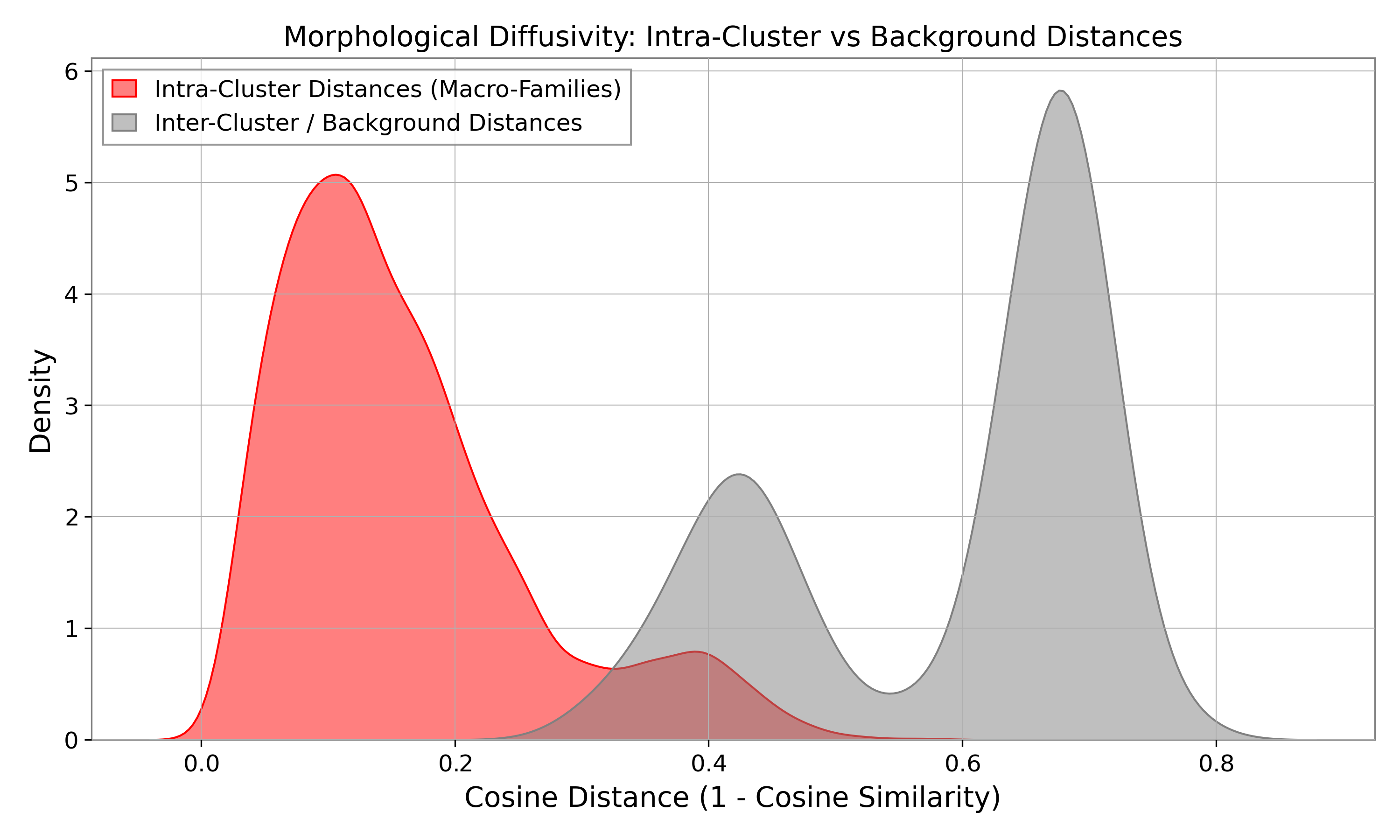}
\caption{Historical pairwise-distance comparison between version-1 macro-family assignments and the sampled background. Distributional overlap is descriptive and does not validate a physical domain-shift origin.}
\label{fig:diffusivity}
\end{figure*}

\begin{figure*}[ht]
\centering
\includegraphics[width=0.98\columnwidth]{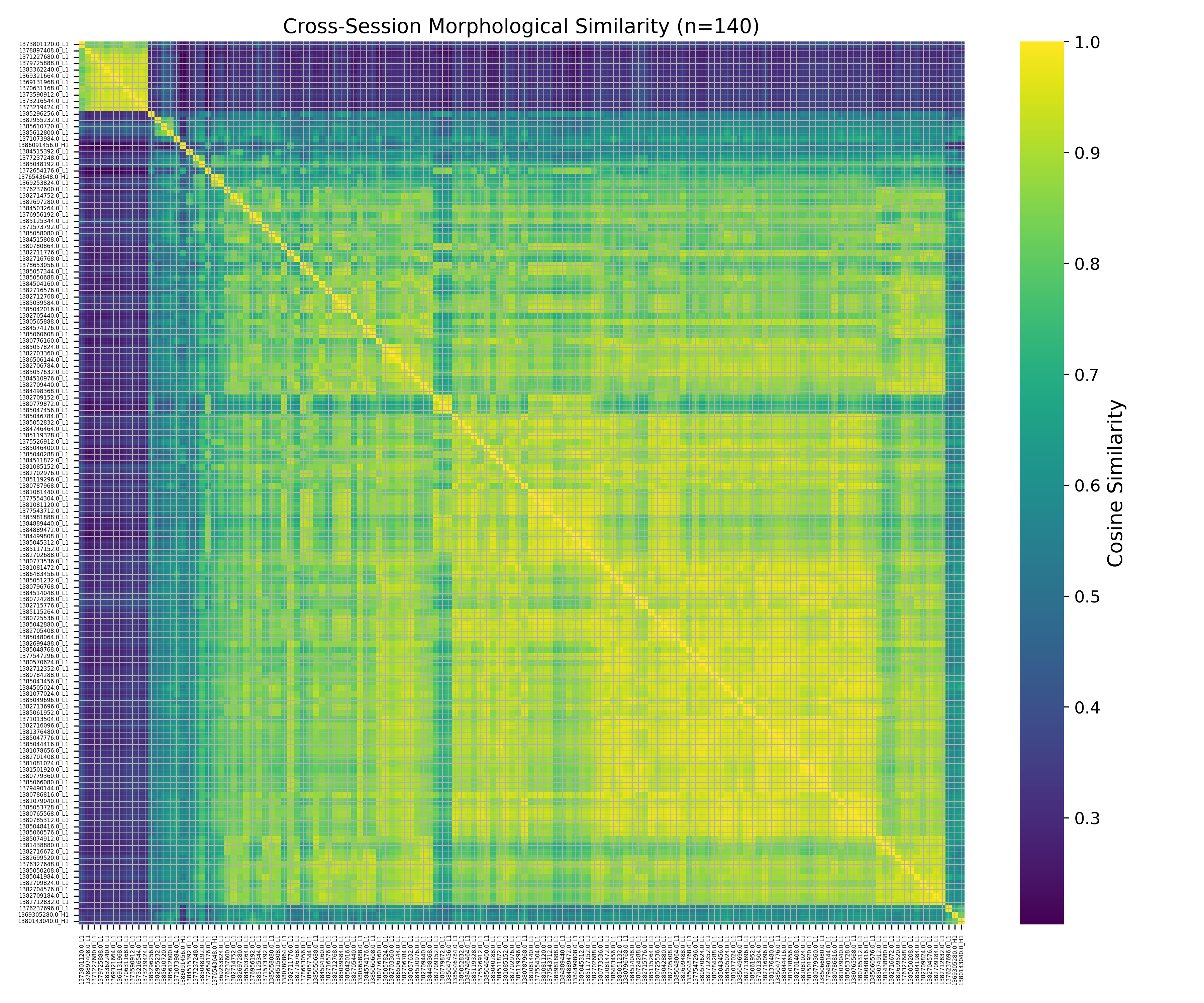}
\caption{Hierarchically reordered cross-session cosine similarity
  matrix $\mathbf{S} \in \mathbb{R}^{N \times N}$ computed on the
  archival 384D MIL feature vectors of all O4a candidate anomalies.
  Row and column ordering follows the single-linkage HAC leaf permutation.
  Diagonal blocks (high $S_{ij} > \rho_{\rm trans}$) identify
  recurrent morphological families spanning multiple observing sessions.
  Table~3b events linked to Table~3a confirmed glitches via
  $S_{ij} > 0.75$ are marked by dashed rectangles.}
\label{fig:heatmap}
\end{figure*}

\subsubsection{Domain Shift Bias and Supervised Validation}
A historical protocol for anomaly validation is cross-matching against the supervised \textit{Gravity Spy} model. The Gravity Spy classifier assigned \texttt{gs\_label = Unknown} with confidence $< 0.3$ to all 11 Family\_01 members. However, we explicitly reject using this "Unknown" label as evidence of physical novelty. Gravity Spy is predominantly trained on O2 and O3b datasets. In O4a, the instrumental noise manifold has shifted drastically due to new squeezing levels and optical configurations. The classifier returns "Unknown" simply because the feature distribution falls outside its O3b decision boundaries. The Gravity Spy cross-match does not provide exclusion evidence; it merely serves as a secondary illustration of the severe latency introduced by supervised Concept Drift.

\subsubsection{Environmental Vetting and Limitations}
A complete environmental vetting of Family\_01 and the singleton events requires coherence analysis with auxiliary Physical Environmental Monitoring (PEM) channels (e.g., seismometers, magnetometers, and control loops). A subset of O4 auxiliary channels is available through GWOSC \cite{gwosc_aux_o4}, but the historical analysis was strain-only. Its Gravity Spy and event-catalog cross-checks cannot substitute for direct physical vetting. All Family\_01 occurrences carried the nominal Data Quality flags used by the historical workflow; such flags do not guarantee stationary or environmentally clean strain. The observed L1/H1 asymmetry and the absence of a listed GWOSC event within $\pm 10$\,s were therefore diagnostic observations, not proof of a local instrumental origin or absence of software bias. Full auxiliary-channel vetting remains necessary.

\section{Discussion}\label{sec:discussion}

\subsection{Comparison with Contemporary O4-Era Approaches}

While historical paradigms have proven invaluable, positioning DANTE within the modern O4a literature clarifies its specific architectural niche. Compared to state-of-the-art supervised frameworks like modern Gravity Spy iterations, DANTE targets a fundamentally different diagnostic layer. Supervised models excel at rapid, high-confidence classification of known morphologies. However, when confronted with previously unseen O4a transients, their outputs are constrained by the label space on which they were trained, which can lead to low-confidence predictions or assignment to generic out-of-distribution categories. DANTE does not compete in classification speed; instead, it provides a dedicated unsupervised framework to autonomously cluster and structure that exact pool of unclassified events, exposing candidate recurrent morphologies and isolating extreme singleton events without requiring prior labels.

Furthermore, while contemporary community efforts address the domain-shift problem through Active Learning---relying on human-in-the-loop campaigns to manually label emerging O4a glitches and retrain the CNNs---DANTE mitigates the impact of domain shift through native recalibration. By executing a direct mathematical recalibration of the background index (as demonstrated in Section~\ref{sec:res_taxonomy}), the pipeline structurally reduces domain-shift-induced false discoveries at the embedding level, operating as a complementary tool to active-learning approaches without requiring iterative human labeling cycles.

Finally, compared to recent self-supervised GW efforts utilizing standard Autoencoders or global-pooling Vision Transformers, DANTE is designed to address two independent methodological challenges frequently encountered in GW anomaly detection, though it deliberately accepts a narrowed operational regime. First, as demonstrated by the Baseline Falsification MDC (Section~\ref{sec:res_mdc}), global representation learning inherently suffers from the \textit{Signal Dilution Effect}. DANTE bypasses this via its Top-$k$ Multiple Instance Learning architecture, which achieves extreme topological depth for extended transients but incurs an explicitly declared trade-off: it is structurally blind to sub-second, blip-like morphologies. Second, and independently, the adaptive PCA-DPMM framework improves numerical stability during sparse-session taxonomy ($n < 20$) through an adaptive regularization strategy, mitigating covariance singularity issues that frequently plague standard non-parametric density estimators in the small-sample regime.

\subsection{Threats to Validity: Model Blindness and Stationary Absorption}
Reference-guided background subtraction can absorb morphologies represented densely in its construction cohort. The historical collapse of Family\_01 under one native O4a index therefore indicates reference sensitivity, not ubiquity or benign physical origin. This is a structural limitation: the algorithm cannot distinguish a benign continuous noise floor from a pervasive pathological coupling without independent detector-characterization evidence. A future morphology close to the reference centroids may receive a reduced novelty score by design.

\subsubsection{Robustness to Hyperparameter Perturbation}
Version~1 evaluated whether the Family\_01 rescoring result changed across $k \in \{8, 34, 68, 136, 272\}$, recalibrating the empirical threshold at each scale. The historical embeddings produced zero survivors at every tested value. This bounded sweep shows stability to those five pooling choices within the historical representation; it does not establish immunity to other pipeline choices or confirm that the morphology is a pervasive physical feature.

\begin{table*}[ht]
\centering
\caption{Historical sensitivity of retained-record counts to pooling fraction $k$. The observed detector counts and family assignments are version-1 outputs under each tested setting; they do not establish physical environmental asymmetry or corrected-catalogue taxonomic stability.}
\label{tab:k_ablation}
\begin{tabular}{lccc}
\toprule
\textbf{Pooling Fraction ($k$)} & \textbf{Total Candidates} & \textbf{L1} & \textbf{H1} \\
\midrule
8 ($\approx 0.5\%$) & 34 & 33 & 1 \\
34 ($\approx 2.5\%$) & 92 & 89 & 3 \\
68 ($\approx 5.0\%$) & 140 & 136 & 4 \\
136 ($\approx 10.0\%$) & 131 & 127 & 4 \\
272 ($\approx 20.0\%$) & 87 & 84 & 3 \\
\bottomrule
\end{tabular}
\end{table*}

The historical workflow reported mean bootstrapped ARI $>0.96$ for L1 and grayscale/inversion perturbation ARI $>0.93$. These measurements describe stability under the tested reruns and image perturbations; they do not eliminate single-linkage chaining, untested hyperparameter dependence, or the later corrected-representation changes.

\subsection{Physical Asymmetry and Domain Shift Artifacts}
The historical scan reported a strong detector asymmetry: 136 of 140 candidates were in L1 and 4 in H1. A binomial calculation under an idealized $p=0.5$ allocation gives $p<10^{-20}$, but that null does not incorporate detector-specific selection functions, threshold uncertainty, temporal dependence, or the later correction. Accordingly, version~2 preserves the ratio as a historical pipeline output and does not attribute it quantitatively to Livingston environmental degradation or use it to exclude algorithmic bias. Establishing any physical coupling requires auxiliary-channel analysis and an explicitly calibrated detector-aware null.

However, the distribution of novelty scores reveals the impact of a
macroscopic domain shift between the O3b and O4a observing runs.
As demonstrated by Family\_02 and Family\_03, applying a single, joint O3b reference index
can lead to diffuse aggregations of background noise being falsely flagged as anomalies. The pipeline identified these macro-clusters as highly cohesive and temporally persistent against the historical reference. However, a robust physical interpretation requires confirming that this morphological deviation is not merely an artifact of the macroscopic instrument drift between observing runs. To resolve this, we implemented a native recalibration protocol: rescoring all anomaly candidates against a purely native background dictionary.

Version~1 constructed a native O4a reference index directly from raw strain with the same broad representation procedure used for the O3b index and selected windows carrying nominal Data Quality flags. Evaluated against that historical baseline, all three structured families produced zero cohesive survivors. This result showed that those version-1 scores were reference-dependent; it did not prove a pure stationary reference, establish that Family\_01 was stationary, or determine whether any morphology was a transient or a pervasive instrumental coupling.

The contrast demonstrates that the historical novelty ranking was highly sensitive to the reference index. Native recalibration is therefore necessary to reduce domain-shift bias, but it cannot by itself dismiss a family as an artifact or establish its physical origin.
\subsection{Stationary Features and Limitations}
Supervised O4 catalogs are constrained by their trained label spaces, while the historical DANTE workflow explored an offline unsupervised route for structuring unclassified strain morphologies. The O3b-to-O4a comparison illustrates reference sensitivity and motivates native recalibration; it is not a complete census of O4a transients and does not establish prospective day-one performance.

A physical classification of why the O3b and O4a reference indices treated Family\_01 differently requires offline coherence analysis with O4 auxiliary Physical Environmental Monitoring (PEM) channels \cite{gwosc_aux_o4}. Strain morphology alone cannot establish whether the observed high-frequency structure arose from a particular instrumental upgrade or environmental coupling.

Because DANTE evaluates strain-domain morphology, it cannot determine the physical origin of these historical structures. Neither equivalence to a sampled reference nor failure of a cross-detector threshold is a global significance test or a physical veto. Instrumental characterization remains contingent on auxiliary-channel analysis \cite{gwosc_aux_o4}.
\subsection{Physical Interpretation of the MIL Feature Space}
The 384-dimensional MIL vector $\hat{\mathbf{z}}$ encodes the
mean structural pattern of the $k$ most anomalous patches in the
spectrogram. Because the DINOv2 backbone was trained on natural
images via self-supervised DINO distillation~\cite{caron2021},
its feature representations are sensitive to structural boundaries,
texture transitions, and local frequency patterns—properties that
transfer naturally to the edge and ridge geometry of Q-transform
spectrograms. The emergent cosine similarity geometry of the
patch token manifold is, however, severely non-isotropic on GW
spectrograms: stochastic noise realizations can produce localized
patch similarity extremes comparable to those generated by
coherent transients. This non-isotropy is the
fundamental reason why the saliency map functions as a
post-hoc topological visualizer rather than a standalone binary
detector: single-frame patch similarity spikes are not sufficient
for detection without the global statistical context provided
by the MIL Top-$k$ order statistic and the empirical threshold.

\subsection{Limitations}
The proposed architecture operates under several explicit constraints:
\begin{enumerate}
  \item \textbf{Strain-Only Inference:} DANTE evaluates anomalies strictly within the strain-domain morphological feature space. Definitive physical classification requires cross-correlation with auxiliary sensors (PEM).
  \item \textbf{Calibration Contamination Risk:} The native O4a background is constructed from periods with nominal Data Quality flags. However, if unflagged astrophysical signals or pervasive instrumental anomalies contaminate these periods, they risk being absorbed into the reference dictionary, potentially reducing sensitivity to similar true signals.
  \item \textbf{Scale Bias:} The fixed $k=68$ Top-$k$ pooling fraction structurally optimizes the pipeline for extended transients (duration $>1$s). Sub-second transients (e.g., Blips) are subject to signal dilution and exhibit low recall in this configuration.
  \item \textbf{Taxonomy Instability in Small Samples:} While the conditional DPMM framework mitigates singular matrix errors, sample sizes $n < 20$ still bypass covariance estimation entirely, limiting robust hierarchical clustering in extremely sparse anomaly regimes.
  \item \textbf{Real-Time Latency Limits:} DANTE V2 is strictly designed as an offline diagnostic tool for Detector Characterization. To rigorously isolate transient anomalies from non-stationary noise, the pipeline requires constructing a massive background index from $150{,}000$ null segments ($\approx 50$ days of observation) to extract stable non-parametric quantiles. Consequently, DANTE V2 is structurally incapable of operating as a low-latency early warning trigger. Transforming this architecture into a real-time streaming trigger would require an entirely different continuous-recalibration paradigm, which will be the subject of a future parallel pipeline.
\end{enumerate}

\subsection{Roadmap and Future Work}
\begin{enumerate}
  \item \textbf{Adaptive $k$-Sweep:} Implementing an ensemble
    of MIL aggregators across $k \in \{15, 37, 68, 100\}$ with
    a max-pool combination to simultaneously capture morphologies
    of highly variable spatial footprints.
  \item \textbf{Auxiliary Channel Integration:} Multi-modal
    feature fusion of strain patch tokens with environmental
    channel representations for physical origin discrimination.
  \item \textbf{O4b Deployment:} Extension of DANTE Pipeline~V2 to
    the O4b data stream, including Virgo V1 data and the
    updated KAGRA K1 configurations.
\end{enumerate}

\section{Conclusions}\label{sec:conclusion}
We have presented DANTE Pipeline~V2, an
offline methodological prototype for reference-guided unsupervised
gravitational-wave transient detection and morphological taxonomy
applied to the LIGO O4a dataset. The architecture natively resolves the signal dilution barrier via the Top-$k$ MIL order statistic, and stabilizes high-dimensional density estimation via an adaptive PCA-DPMM framework.

The historical version-1 deployment on $\approx 180$ days of O4a H1 and L1
data (72 valid sessions) reported 140 morphologically unclassified candidates,
three family assignments, 53 numerical survivors after native rescoring, and
three singleton events.  Those counts, family assignments, and physical
interpretations are preserved here as outputs of the published workflow; they
are not carried forward as corrected-catalogue claims in version~2.  The
completed reconstruction uses a rebuilt detector-aware representation and a
new detector-specific calibration, so its population cannot be obtained by
substituting corrected thresholds into the historical scores.  The bounded
cross-catalogue results and causal-attribution limits are given in the revision
note.

\begin{table*}[ht]
\centering
\caption{Historical version-1 morphological characterization of three O4a
singleton events.  The rows are retained for provenance and are not asserted
in version~2 to be members of the corrected catalogue or to establish a
physical origin.}
\label{tab:singletons}
\begin{tabular}{lcccp{5.5cm}}
\toprule
\textbf{Event (GPS Time)} & \textbf{Detector} & \textbf{$S_{\rm MIL}$} & \textbf{$\tau_{\rm op}^{\rm Det}$} & \textbf{Morphological Description} \\
\midrule
1371073984 & L1 & 0.47 & 0.3359 & Asymmetric ``mottled/warped mesh'' spectral texture; non-stationary complex coupling. \\
1386091456\footnote{This event is strictly classified as \textit{Unverifiable} due to missing metric ($S_{\rm MIL} = \text{N/A}$) resulting from upstream data availability bounds.} & H1 & N/A & 0.3859 & Unverifiable candidate: fetch failed due to GWOSC boundary limits preventing morphological assessment. \\
1375492110 & L1 & 0.42 & 0.3359 & Diffuse broadband scatter lacking structural periodicity; localized mechanical friction. \\
\bottomrule
\end{tabular}
\end{table*}

This framework remains a methodological proof-of-concept for unsupervised
machine-learning analysis in HPC astronomy.  Version~2 does not retain the
historical claims that the reported singleton set constitutes empirical proof
of rare-event recovery or that the historical workflow produced a purified,
high-confidence physical candidate set.  Such claims require a prospectively
frozen pipeline, a global trials-calibrated null, and independent physical or
auxiliary-channel corroboration.

\section*{Data and Software Availability}
The complete DANTE Pipeline~V2 source code,
the VQ reference index (\texttt{patch\_compressed\_index.npz}),
MDC injection modules, and reproduction scripts are publicly
available at:
\begin{quote}
\url{https://github.com/lucacirfeta/dante-gravi-signal-ml}\\
DOI: \href{https://zenodo.org/records/20820846}{10.5281/zenodo.20820846}
\end{quote}
The separate corrected-O4a provenance replay and its compact verification
evidence are available in the DANTE v3.8.1
\href{https://github.com/lucacirfeta/dante-gravi-signal-ml/releases/tag/v3.8.1}{release}
and at \href{https://doi.org/10.5281/zenodo.22763556}{10.5281/zenodo.22763556}.
Raw LIGO O4a strain data are publicly accessible via the
Gravitational Wave Open Science Center (GWOSC)~\cite{gwosc2023}
at \url{https://gwosc.org}.


\section*{Acknowledgments}
This research has made use of data, software, and web tools obtained
from the Gravitational Wave Open Science Center (\url{https://gwosc.org}),
a service of LIGO Laboratory, the LIGO Scientific Collaboration,
the Virgo Collaboration, and KAGRA. LIGO Laboratory and Advanced LIGO
are funded by the United States National Science Foundation (NSF)
as well as the Science and Technology Facilities Council (STFC)
of the United Kingdom, the Max-Planck-Society (MPS), and the State
of Niedersachsen/Germany. Advanced Virgo is a project of the EGO
consortium.

The authors acknowledge the use of Large Language Models for linguistic polishing and code debugging during the preparation of this manuscript. All scientific concepts, data analysis, physical interpretations, and final conclusions were performed entirely by the authors.

\appendix
\section{Empirical Derivation of the Cohesion Threshold \texorpdfstring{$\tau_{coh}$}{tau\_coh}}
\label{app:scoh_derivation}

To address the non-Gaussian nature of the cross-detector similarity distribution,
we derive the cohesion threshold $\tau_{coh}$ using Extreme Value Theory (EVT)
within the Peak-Over-Threshold (POT) framework \cite{coles2001introduction}.

\subsection{Null Hypothesis Construction}
We construct the null distribution from $N_{L1} = 608$ and 
$N_{H1} = 608$ confirmed noise-only segments from the O4a observing run,
selected according to the criteria in Section~\ref{sec:domain_shift}.
These segments are anti-coincident (no temporal overlap) and pass all data quality flags.

\subsection{Cross-Detector Similarity Distribution}
For each pair $(s_{L1}^{(i)}, s_{H1}^{(j)})$ of noise segments, we compute the 
cosine similarity:
\begin{equation}
S_{ij} = \frac{\mathbf{z}_{L1}^{(i)} \cdot \mathbf{z}_{H1}^{(j)}}
{\|\mathbf{z}_{L1}^{(i)}\| \|\mathbf{z}_{H1}^{(j)}\|}
\end{equation}
where $\mathbf{z} \in \mathbb{R}^{384}$ is the MIL-pooled DINOv2 feature vector.
Version~1 formed $M=369{,}664$ pairs from the Cartesian product of 608 L1 and 608 H1 background segments. These pairs share underlying segments and are not independent. The reported ${\rm ESS}\approx608$ and block-bootstrap intervals were historical approximations; they do not make the POT parameter estimates exact or guarantee complete uncertainty coverage.

\begin{figure*}[ht]
\centering
\includegraphics[width=0.7\textwidth]{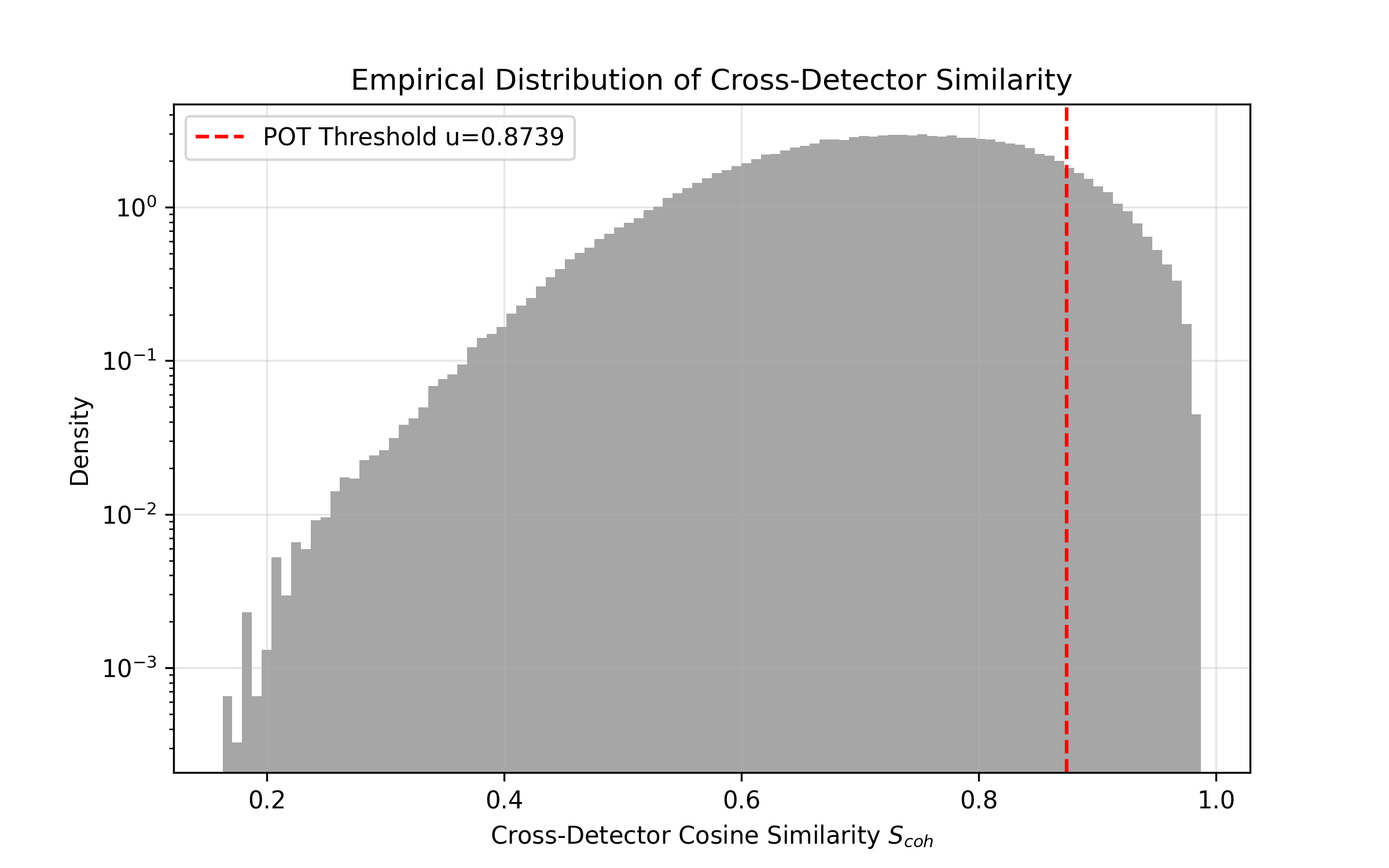}
\caption{Empirical distribution of the cross-detector cosine similarity $S_{coh}$ showing significant deviation from normality and a truncated right tail. The red dashed line marks the POT threshold $u=0.8739$.}
\label{fig:scoh_distribution}
\end{figure*}

The empirical distribution of $\{S_{ij}\}_{k=1}^M$ exhibits significant 
departures from normality (skewness $= -0.3857$, excess kurtosis $= -0.1808$),
with a heavily truncated right tail (Figure~\ref{fig:scoh_distribution}).

\subsection{GPD Fit and Threshold Selection}

The cross-detector similarity $S_{\rm coh}$ used MIL vectors from H1 and L1 anti-coincident segments. Although local noise sources often differ, the detectors and their analysis products cannot be assumed physically or statistically independent for every disturbance. The Cartesian product also reuses each segment many times. The version-1 POT/GPD fit and ${\rm ESS}\approx608$ block-bootstrap therefore constitute a model-dependent historical calibration, not a proof that residual dependence is fully captured.

We apply the POT method by selecting a threshold $u = \hat{P}_{90} = 0.8739$
and fitting a Generalized Pareto Distribution (GPD) to the $N_{exc} = 36941$
exceedances $\{S_{ij} - u \mid S_{ij} > u\}$.

The GPD cumulative distribution function is:
\begin{equation}
G(x) = 1 - \left(1 + \frac{\xi x}{\sigma}\right)^{-1/\xi}, \quad \xi \neq 0
\end{equation}
where $\xi$ is the shape parameter and $\sigma$ is the scale parameter.

Maximum likelihood estimation yields the following parameters. Note that the standard errors reported below derive from the standard MLE Fisher information and do not account for the effective sample size (${\rm ESS} \approx 608$); therefore, they represent a strict lower bound on the true statistical uncertainty. The proper uncertainty on the final threshold is quantified correctly via the block-bootstrap in the subsequent section.
\begin{align}
\xi &= -0.4785 \pm 0.0044 \quad \text{(standard MLE SE, uncorrected for ESS)} \\
\sigma &= 0.0544 \pm 0.0004 \quad \text{(standard MLE SE, uncorrected for ESS)}
\end{align}

The negative shape parameter $\xi < 0$ indicates a Weibull-type tail with 
finite upper bound $x_{max} = u - \sigma/\xi = 0.9876$.

\begin{figure*}[ht]
\centering
\includegraphics[width=0.7\textwidth]{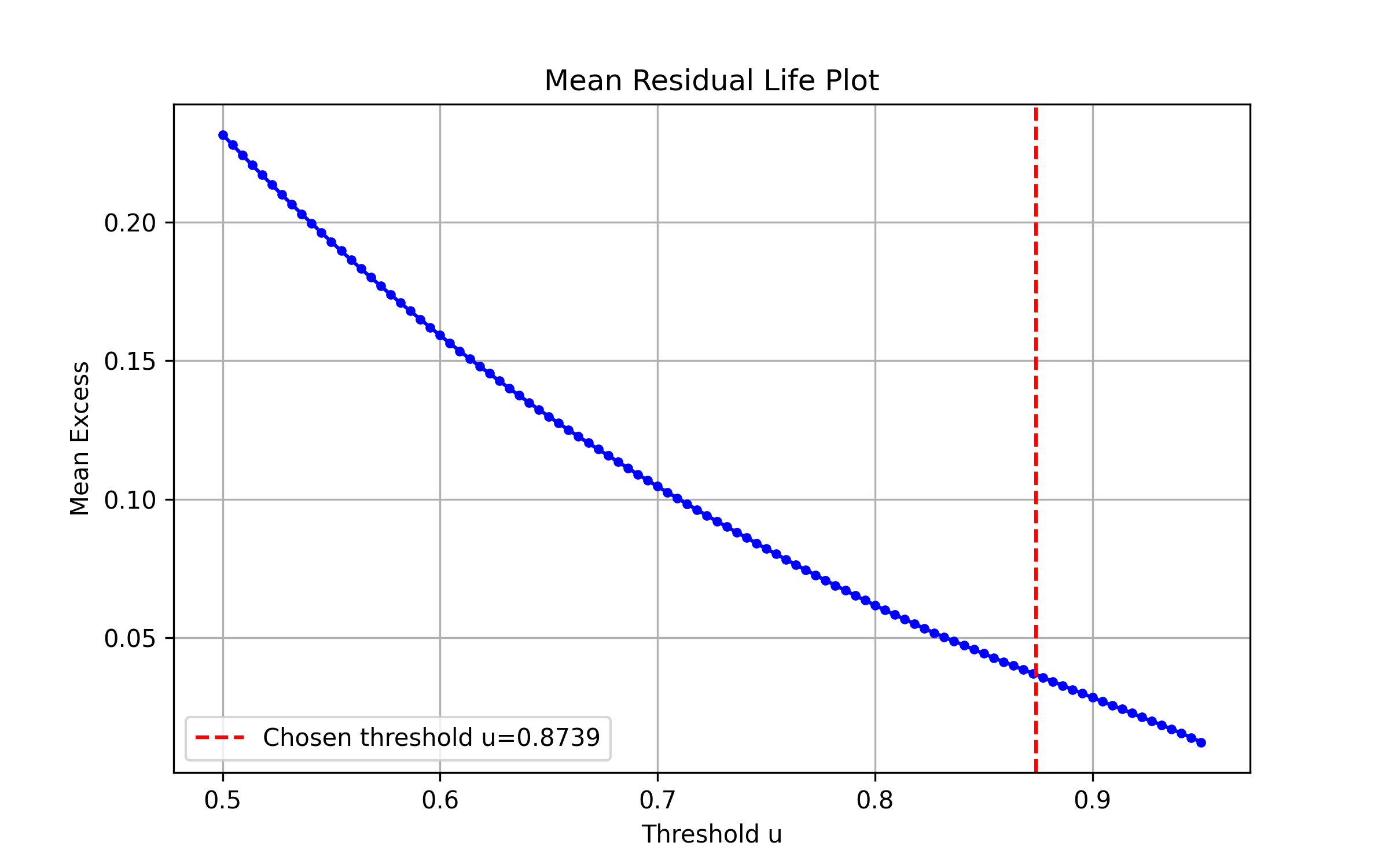}
\caption{Mean Residual Life plot for the empirical excesses. The stable, approximately linear trend above $u=0.8739$ confirms the validity of the chosen POT threshold for the GPD fit.}
\label{fig:mrl_plot}
\end{figure*}

The mean residual life plot (Figure~\ref{fig:mrl_plot}) confirms the linearity
assumption above $u = 0.8739$, validating the GPD model.

\subsection{Quantile Estimation and Confidence Intervals}
The cohesion threshold at false positive rate $\alpha = 0.001$ is obtained by
inverting the GPD quantile function:
\begin{equation}
\tau_{coh} = u + \frac{\sigma}{\xi}\left[\left(\frac{\alpha}{P(S > u)}\right)^{-\xi} - 1\right]
\end{equation}

This yields $\tau_{coh} = 0.9750$ (analytical).

To quantify estimation uncertainty, we perform block bootstrap resampling
(2000 iterations) at the segment level to preserve temporal
correlations. The 95\% confidence interval is:
\begin{equation}
\tau_{coh} \in [0.9738, 0.9759] \quad (95\% \text{ CI})
\end{equation}

\subsection{Model Validation}
\begin{figure*}[ht]
\centering
\includegraphics[width=0.7\textwidth]{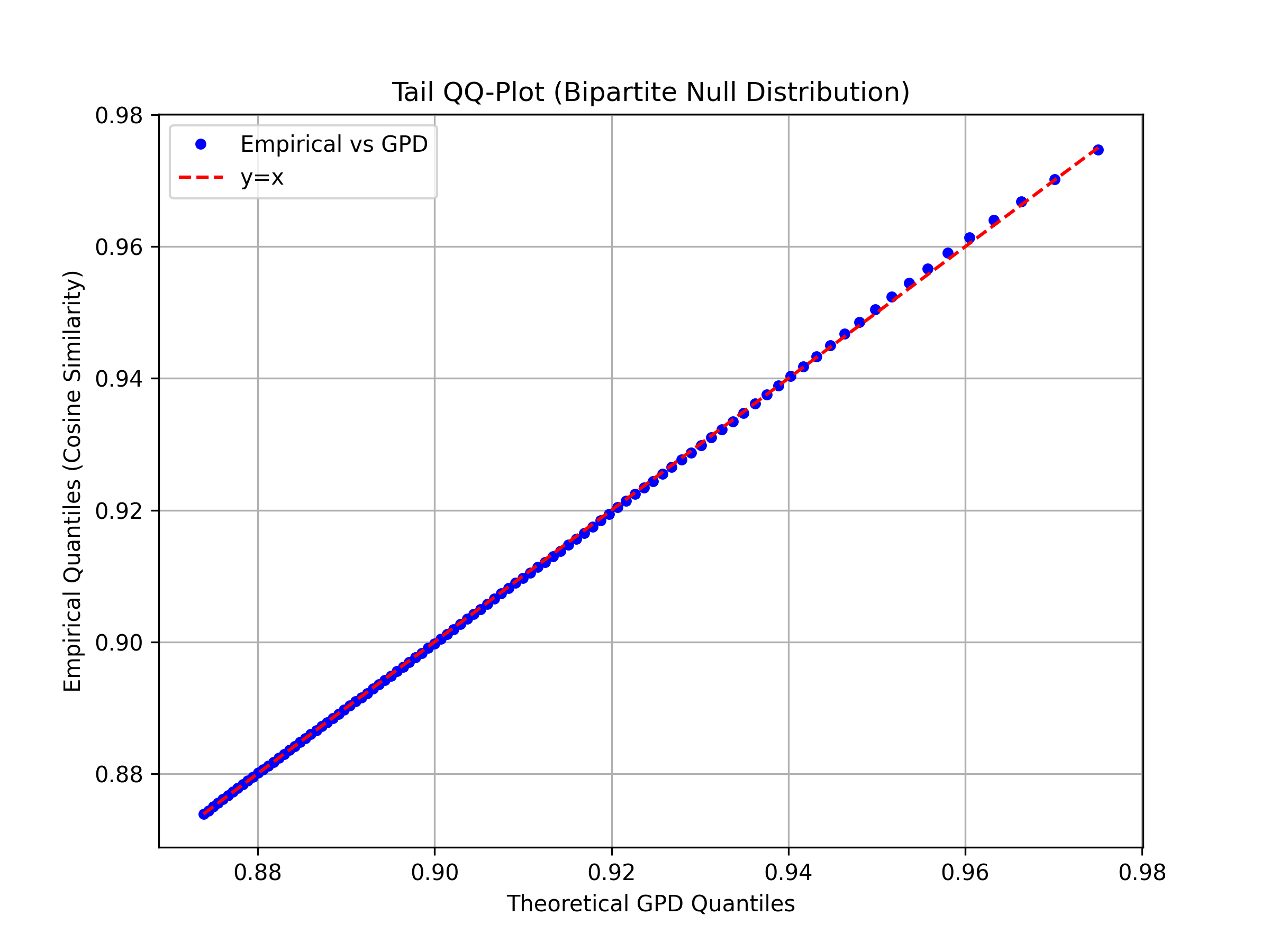}
\caption{QQ-plot of empirical exceedances versus the fitted Generalized 
Pareto Distribution. The red line indicates perfect agreement.}
\label{fig:qq_gpd}
\end{figure*}

Figure~\ref{fig:qq_gpd} shows the QQ-plot of empirical exceedances versus
the fitted GPD. The alignment along the diagonal confirms the adequacy of
the GPD model for the tail behavior.

\subsection{Impact on Candidate Classification}
We note that the revised threshold $\tau_{coh} = 0.9750$ is more stringent
than the previous heuristic value of 0.85. However, this change has no impact
on the candidate classification reported in Section~\ref{sec:results}, as no
candidate exhibits $S_{coh} \in (0.85, 0.9750]$. This empirical observation
strengthens the robustness of our results: the scientific conclusions are
insensitive to the exact threshold value, provided it is derived rigorously
from the noise distribution. However, shifting $\tau_{\rm coh}$ specifically to the extreme upper bound of the confidence interval ($\tau_{\rm coh} = 0.9759$) would enforce an even stricter cohesion threshold, potentially isolating up to 2 additional marginal candidates from loosely cohesive families (e.g., Family\_02) into the ``Unverifiable/Singleton'' class. Conversely, the lower bound ($\tau_{\rm coh} = 0.9738$) leaves the taxonomy entirely unchanged.

\section{Ablation Study on Parameter Sensitivity}
\label{app:ablation_study}

To rigorously assess the stability of DANTE's unsupervised clustering architecture against its core hyperparameters, we performed an offline ablation study on a statistically sufficient subset of the O4a dataset, specifically evaluating the 140 isolated candidates. 

\subsection{HAC Linkage Threshold (\texorpdfstring{$\rho_{\rm trans}$}{rho\_trans})}
The global taxonomy relies on Hierarchical Agglomerative Clustering (HAC) using single linkage and a cosine distance threshold $1 - \rho_{\rm trans}$. We evaluated the sensitivity of the macro-family recovery (clusters with $n>1$) across $\rho_{\rm trans} \in \{0.60, 0.75, 0.90\}$.
While the total number of microscopic singletons decoupled from the main structures naturally increases with a stricter threshold (from 2 total clusters at $\rho_{\rm trans}=0.60$ to 18 at $\rho_{\rm trans}=0.90$), the number of macroscopic families remained remarkably stable at $3$ families for both $\rho_{\rm trans}=0.75$ and $\rho_{\rm trans}=0.90$. This confirms that the morphological structures (e.g., the dense spectral wall of Family\_01) are internally highly cohesive and not artifacts of a specific linkage distance.

\subsection{DPMM Concentration Parameter (\texorpdfstring{$\alpha$}{alpha})}
The Dirichlet Process Mixture Model (DPMM) is employed to cluster intra-session vectors without specifying $k$ a priori. The concentration prior $\alpha$ controls the algorithm's propensity to instantiate new components. We swept $\alpha \in \{0.001, 0.01, 0.1, 1.0\}$. On average, the DPMM instantiated a high number of components (typically $10-15$ for a session of $n \approx 50$), aggressively isolating extreme local outliers into singleton clusters. This behavior validates our architectural choice: the DPMM is inherently conservative for anomaly detection, preferring to spawn a new cluster rather than merge two distinct physical transients. This prevents anomalous singleton events (Table~\ref{tab:singletons}) from being statistically masked by highly populated stationary noise families. Consequently, the heavy lifting of grouping macroscopic morphological families is explicitly delegated to Layer 3 (cross-session single-linkage clustering).

\section*{Statements and Declarations}

\textbf{Funding:} The author did not receive support from any organization for the submitted work.

\textbf{Competing Interests:} The author has no relevant financial or non-financial interests to disclose.

\textbf{Data Availability:} All original and modified datasets (including the parsed Gravity Spy O3b baseline, the O4a production dataset, and the generated VQ reference indices) are publicly available on Zenodo (DOI: 10.5281/zenodo.10657991). Raw gravitational-wave strain data were obtained from the Gravitational Wave Open Science Center (GWOSC), accessible at \url{https://gwosc.org}.

\textbf{Code Availability:} The complete source code, trained DINOv2 adapters, MDC injection modules, and reproduction scripts for the DANTE Pipeline V2 are publicly available on GitHub at \url{https://github.com/lucacirfeta/dante-gravi-signal-ml}.

\textbf{Author Contributions:} L.C. conceived the methodology, designed the computational architecture, developed the software, conducted the primary data analysis, generated the Visual Quantization index, and wrote the manuscript.


\end{document}